# Water-stable MOFs and Hydrophobically Encapsulated MOFs for CO$_2$ Capture from Ambient Air and Wet Flue Gas

Xiaoyang Shi,[1,2,3,‡] Gahyun Annie Lee[2,3,‡], Shuohan Liu[4], Dongjae Kim[1,3], Ammar Alahmed[5], Aqil Jamal[5], Lei Wang[*4], Ah-Hyung Alissa Park[*1,2,3]

[1] Department of Earth and Environmental Engineering, Columbia University, New York, NY 10027, USA

[2] Department of Chemical Engineering, Columbia University, New York, NY 10027, USA

[3] Lenfest Center for Sustainable Energy, The Earth Institute, Columbia University, New York, NY 10027, USA

[4] National Laboratory of Solid-State Microstructures, School of Physics and Collaborative Innovation Center of Advanced Microstructures, Nanjing University, Nanjing 210093, China

[5] Research and Development Center, Saudi Aramco, Dhahran 31311, Saudi Arabia

‡ These authors contributed equally

*Corresponding authors: ap2622@columbia.edu, Leiwang@nju.edu.cn

## Abstract:

The extra CO$_2$ that has already been released into the atmosphere has to be removed in order to create a world that is carbon neutral. Technologies have been created to remove carbon dioxide from wet flue gas or even directly from ambient air, however these technologies are not widely deployed yet. New generations of creative CO$_2$ capture sorbents have been produced as a consequence of recent improvements in material assembly and surface chemistry. We summarize recent progress on water-stable and encapsulated metal-organic frameworks (MOFs) for CO$_2$ capture under a wide range of environmental and operating conditions. In particular, newly developed water-stable MOFs and hydrophobic coating technologies are discussed with insights into their materials discovery and the synergistic effects between different components of these hybrid sorbent systems. The future perspectives and directions of water-stable and encapsulated MOFs are also given for Direct Air Capture of CO$_2$ and CO$_2$ capture from wet flue gas.

# 1. Introduction:

The main cause of global warming is the buildup of carbon dioxide ($CO_2$) in the atmosphere[1]. The Intergovernmental Panel on Climate Change (IPCC) scenario RCP 8.5 states that, in the absence of action, $CO_2$ emissions would increase from the current level of 49 GtCO2eq/yr to between 85 and 136 GtCO2eq/yr by the year 2050[2]. Global mean temperatures might rise by 3.8 to 6.0 °C from their pre-industrial level (1880–1900) to 2100 as a result of growing $CO_2$ concentrations. While supplying the increasing energy demand, $CO_2$ emissions might be reduced through $CO_2$ collection from wet flue gas mixes produced by industrial producers and power plants, or from the ambient air directly[3-6].

Metal-organic frameworks (MOFs) are three-dimensional (3D) materials made of metal ions or clusters connected by organic ligands (linkers). MOFs exhibit high $CO_2$ capacity and selectivity for $CO_2$[7] due to their high surface functionality and porosity[8-11], and highly unusual tunability[12-13]. It has been demonstrated that some metal-organic frameworks' separation capacities are on par with or better than those of zeolite or carbon adsorbents[7, 14-16]. However, several investigations show that MOFs have serious stability problems[17-22], especially concerning their interactions with water[23-24], severely limiting their practical potential. Comparing the usual chemical properties of post-exposure and pristine samples will reveal if a MOFs structure is still stable in the water stability test. Indicating whether the MOFs loses its crystallinity or structural porosity following exposure to moisture or humid streams are the chemical properties of the powder X-ray diffraction (PXRD) pattern and BET surface area based on gas adsorption capability. MOFs structures are often vulnerable to hydrolysis, which might result in ligand displacement, phase shifts, and structural breakdown.

When evaluating sorbents for $CO_2$ collection, it is important to take into account the presence of water vapor, which is present in ambient air (0.2 percent–4 percent) and industrial flue gases (10 percent). Increasing MOFs stability has been a major objective of current research[25-37]. In order to strengthen the bonding in the material, most of this research has focused on altering the chemical makeup of the MOFs itself (either the metal node or the organic linker). Using high-oxidation metal ions, like $Zr^{4+}$ in MOFs like UiO-66[25], or changing the linker units with other functional groups, such pyrazolate[26], both boost the metal-linker bond strengths, are successful examples in this field. Other techniques involve post-synthetic changes to shield the MOFs from water intrusion, such as hydrophobic coatings or mixed-matrix composites[38-40]. These methods alter the chemistry of the substance while boosting MOFs stability. Such chemical alterations frequently cause a diminution in the solid's functioning, sometimes gravely jeopardizing its practicability[17]. As a result, there is a critical need to create MOFs with the ideal chemical structure for $CO_2$ collection and sufficient hydrolytic stability for practical use. This paper examines techniques that have been suggested to be effective for $CO_2$ collection in humid environments and gives a general overview of the factors to take into account when coming up with a fresh strategy to increase MOFs stability in humid environments.

## 2. Design and synthesis of MOFs

Strong coordination bonds (thermodynamic stability) or considerable steric hindrance (kinetic stability) are typically present in MOFs structures with outstanding stability to prevent the damaging hydrolysis process that destroys the metal-ligand connections[17]. Numerous research articles on water-stable MOFs are now witnessing a rise, and many more water-stable MOFs are published every year thanks to ongoing efforts and greater understanding of MOFs structural stability in humid surrounding. Due to the important work done by Burtch *et al.*[17], Canivet *et al.*[41], Howarth *et al.*[42] *etc,* a comprehensive database of water-stable MOFs has so far been created. Direct synthesized water-stable MOFs may be divided into three main groups: (1) High-valence metal ion-based metal carboxylate frameworks; (2) Metal azolate frameworks with nitrogen-donor ligands; and (3) MOFs with hydrophobic pore surfaces or metal centers that are not exposed. Methods for stabilization of MOFs in water according to direct synthesis are shown in Figure 1.

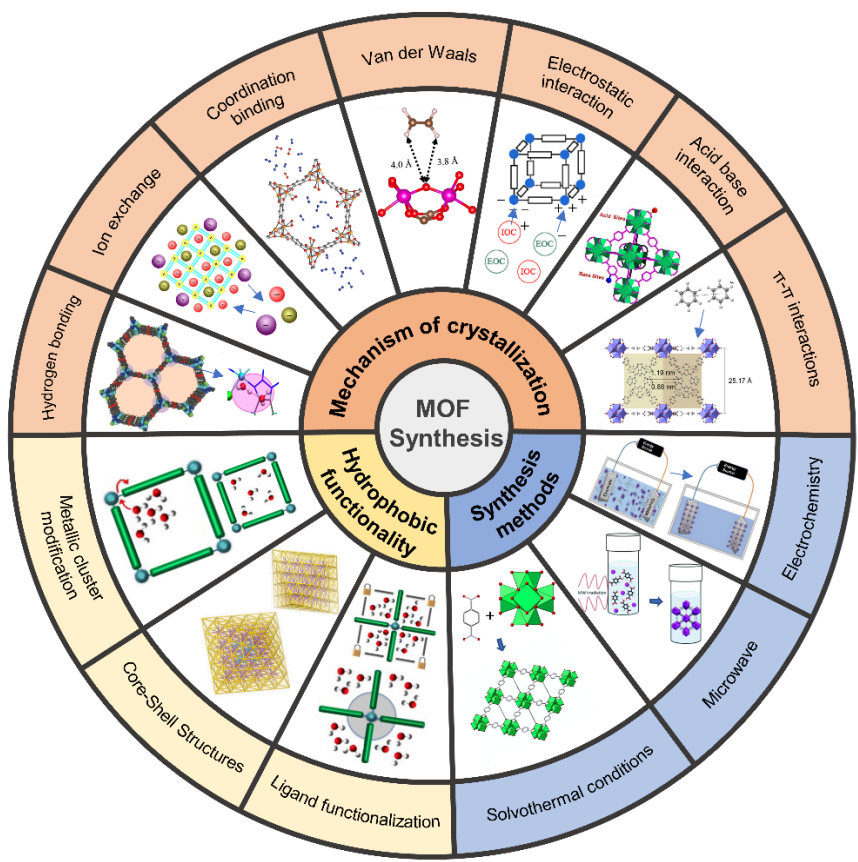

Figure 1. Mechanisms, reaction conditions and incorporation of hydrophobic functionality of MOF synthesis.

## 2.1 Metal Carboxylate Frameworks

High-valence metal ions including $Fe^{3+}$, $Cr^{3+}$, and $Zr^{4+}$ have been employed as common carboxylate-type ligands to create water-stable MOFs. Due to their great aqueous stability, the combination of aromatic carboxylate molecules with a cluster caused steric hindrance in MOFs, which resulted in a significant depassivation of $H^+$ (or $OH^-$) ion activity on its surface[43]. For instance, Ferey and his colleagues created the renowned Cr-based MIL-101, which offered reasonable chemical stability and could withstand multiple solvents and ambient conditions for months[20]. These characteristics make MIL-101 an appealing choice for the adsorption of gas, in addition to its high adsorption capabilities. At 78 K, the MIL-101 nitrogen gas sorption isotherm is approximately 1200 $cm^3/g$, or 5900 $m^2/g$.

Additionally, high-valence $Zr^{4+}$ cation-containing MOFs such as the well-known UiO-66 and the PCN family, which includes PCN-222, PCN-223, and PCN-224, exhibit extraordinary hydro-stability even in the presence of water[25, 44]. The Zr MOFs structures made with linear ligands are shown in Figure 2A. The structural resilience to various solvents was also examined. A desolvated sample was agitated at room temperature in various solvents for 24 hours. A modest quantity of Si was added to each sample in order to examine alterations in the unit cell parameters. Figure 2B demonstrates the results achieved.

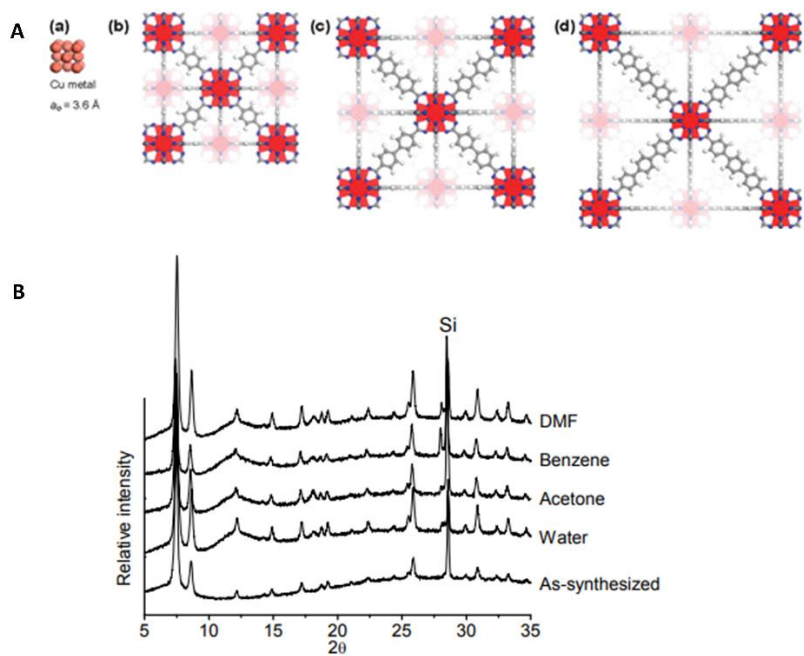

Figure 2 **A.** (a) One unit cell of copper drawn to scale with: (b) Zr−MOFs with 1,4-benzene-dicarboxylate (BDC) as linker, UiO-66, (c) Zr−MOFs with 4,4′ biphenyl-dicarboxylate (BPDC) as linker, UiO-67, (d) Zr−MOFs with terphenyl dicarboxylate (TPDC) as linker, UiO-68. Zirconium, oxygen, carbon, and hydrogen atoms are red, blue, gray, and white, respectively. **B.** Powder XRD patterns of assynthesized UiO66 and UiO66 treated with different solvents following desolvation[22].

In recent years, a variety of water-stable MOFs have been synthesized and described using this method[45-64]. Le *et al.*[65], for instance, describe the utilization of water-stable lanthanide (Ln)-based MOFs as catalysts for converting $CO_2$ into products with added value. The authors discuss the fixing of carbon dioxide into the epoxy ring of propylene oxide for the creation of cyclic carbonates using these MOFs.

## 2.2 Metal Azolate Frameworks

Utilizing azolate ligands (such as imidazolates, pyrazolates, triazolates, and tetrazolates) is an additional technique for the synthesis of water-stable MOFs, in addition to the use of high-valence metals as hard acids[66]. As these nitrogen-containing ligands are typically softer, they can interact with the softer divalent metal ions to produce more robust MOFs structures. The establishment of a strong coordination connection between organic ligands and central metal ions may best explain this observation.

The best illustration of this group is zeolitic imidazolate frameworks (ZIFs). Using $Zn^{2+}/Co^{2+}$ and imidazolate linkers, scientists fabricated a variety of stable crystals with zeolite-like structure[67-69]. The chemical stability of the samples was evaluated by heating them in different solvents for seven days; these circumstances match probable harsh industrial needs. Several ZIFs with heterolinks have a high porosity that can impede the collection and storage of $CO_2$ with outstanding selectivity. Figure 3 depicts the chemical structures of ZIF-68, 69, and 70, as well as their gas adsorption isotherms and $CO_2$ capture characteristics. The Langmuir surface areas for ZIF-68, 69, and 70 were 1220, 1070, and 1970 $m^2\ g^{-1}$, respectively.

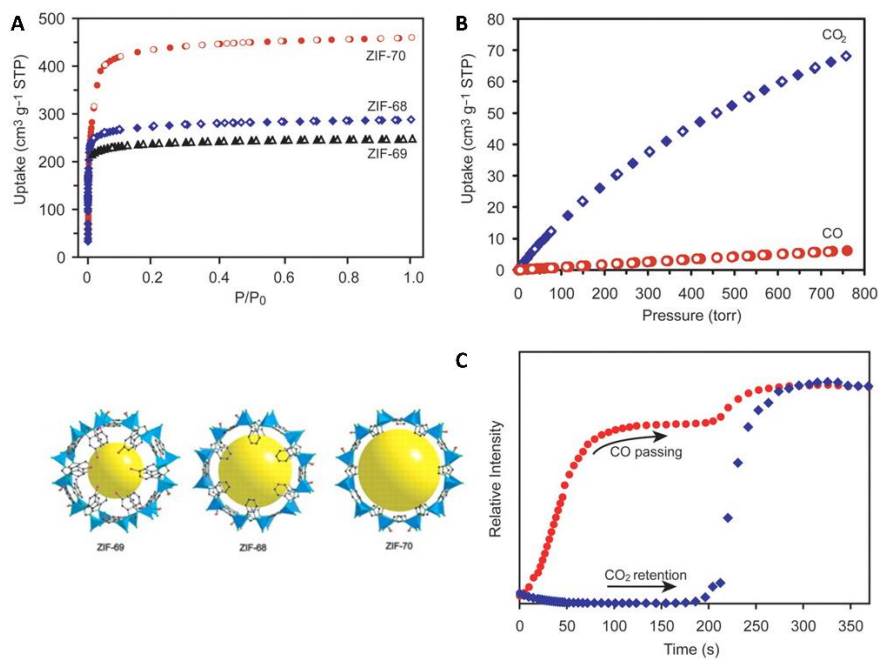

Figure 3. Isotherms of gas absorption and $CO_2$ capture characteristics of ZIFs. **A.** The $N_2$ adsorption isotherms for heterolinked ZIF-68, 69, and 70 at 77 K. $P/P_0$, relative pressure; STP, standard temperature and pressure. **B.** The $CO_2$ and CO adsorption isotherms for ZIF-69 at 273 K. For **A** and **B**, the gas uptake and release are indicated by solid and open symbols, respectively. **C.** Breakthrough curves of a stream of $CO_2$/CO combination that went through a sample of ZIF-68 demonstrate the retention of $CO_2$ in the pores and passage of CO[67].

In addition, Colombo et al.[26] produced the microporous pyrazolate-based MOFs, $M_3(BTP)_2$ (M = Ni, Cu, Zn, Co), which demonstrated superior hydrothermal stability compared to most carboxylate-based MOFs. As a result of this approach, an increasing number of azolate-based MOFs exhibit a respectable level of hydro-stability that was produced in recent years[70-88].

In order to produce a flexible porous coordination polymer with a high $CO_2$ adsorption enthalpy and $CO_2/N_2$ selectivity, Liao et al.[71] functionalized it with two uncoordinated triazolate N-donor pairs. Metal Azolate Framework (MAF-23) is seen in Figure 4B at 273 and 298 K, where it absorbs $CO_2$ at a rate of 74.2 and 56.1 $cm^3$ $g^{-1}$ (14.6 and 11.0 wt percent, respectively) at 1 atm. MAF-23 only adsorbed 4.0 and 2.0 $cm^3$ $g^{-1}$ of $N_2$ at 1 atm under the identical circumstances. Figure 4C demonstrates that even after refluxing in water for seven days, MAF-23 entirely maintained its crystallinity.

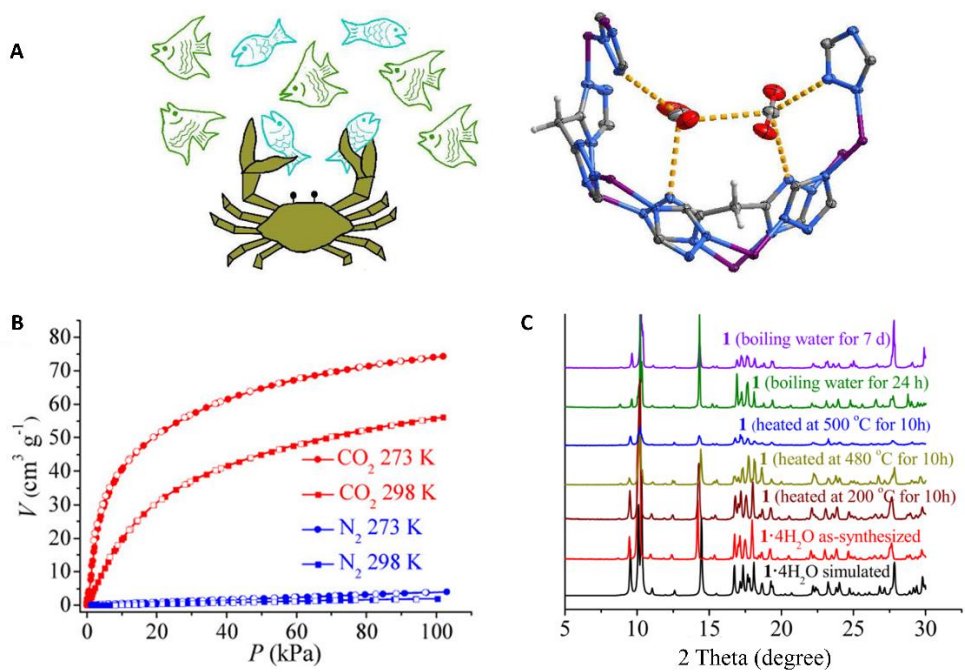

Figure 4**A.** Framework Structure of metal azolate. **B**. $CO_2$ and $N_2$ adsorption (solid) and desorption (open) isotherms. C. PXRD patterns of MAF-23[71].

The bulk synthesis of high-quality MAF-6 with excellent purity, crystallinity, and thermal/chemical stability was accomplished by He *et al.*[74]. When heated at 400 °C in a nitrogen environment for one hour or submerged in methanol, benzene, and water at ambient temperature for at least three days, the original crystallinity could still be maintained, according to PXRD. Figure 5A demonstrates that the surface of MAF-6 is very hydrophobic with a contact angle of 143 ± 1°. At 77K, the $N_2$ sorption isotherm of MAF-6 revealed a Langmuir surface area of 1695 m$^2$/g, a Brunauer–Emmett–Teller surface area of 1343 m$^2$ g$^{-1}$, as well as a pore volume of 0.61 cm$^3$/g (Figure 5B).

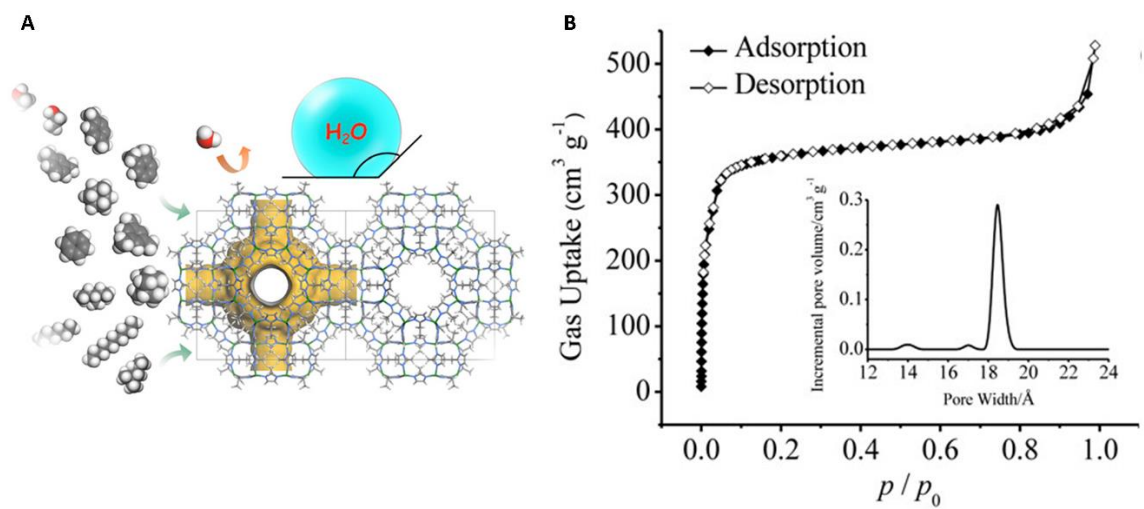

Figure 5. **A**. Hydrophobic structure of metal azolate framework. **B**. $CO_2$ and $N_2$ adsorption (solid) and desorption (open) isotherms[74].

Zhang *et al.*[79] described a coligand-bridging approach for manufacturing novel poly MOFs. These poly MOFs materials displayed reasonably strong $CO_2$ sorption but extremely low $N_2$ sorption, making them interesting for $CO_2$/$N_2$ separations. In addition, the hydrophobicity of polymer ligands and the cross-linking of the polymer chains inside the MOFs contributed to the excellent water stability of these poly MOFs, that were immersed in water at room temperature or 100 °C for 1 day. Five days of solvent exchange with methanol were followed by ten hours of activation at 130 °C under vacuum. PXRD indicated that, with the exception of Zn-pbdc-7a(bpe), all other poly MOFs kept their crystallinity following water treatment. According to the $CO_2$ sorption results reported in Table 1, Zn-pbdc-11a(bpe), Zn-pbdc-12a(bpe), and Zn-pbdc-12a(bpy) demonstrated highly comparable $CO_2$ uptake values before and after water treatment, indicating outstanding water stability. Other poly MOFs materials exposed to water lost some porosity.

| polyMOFs | contact angle (deg) | $CO_2$ uptake $(cm^3/g)^a$ | $CO_2$ uptake $(cm^3/g)^b$ |
|---|---|---|---|
| Zn-pbdc-8a(bpy) | 0 | 91 ± 5 | 70 ± 5 |
| Zn-pbdc-9a(bpy) | 0 | 86 ± 3 | 72 ± 4 |
| Zn-pbdc-12a(bpy) | 119 ± 1 | 75 ± 5 | 78 ± 3 |
| Zn-pbdc-7a(bpe) | 0 | 72 ± 2 | 40 ± 7 |
| Zn-pbdc-8a(bpe) | 0 | 97 ± 5 | 70 ± 1 |
| Zn-pbdc-9a(bpe) | 0 | 80 ± 4 | 61 ± 5 |
| Zn-pbdc-10a(bpe) | 111 ± 1 | 140 ± 5 | 102 ± 6 |
| Zn-pbdc-11a(bpe) | 114 ± 1 | 106 ± 5 | 101 ± 5 |
| Zn-pbdc-12a(bpe) | 115 ± 1 | 105 ± 6 | 106 ± 5 |
| MOF 1 | N/A | 156 | N/A |
| MOF 1' | 0 | 160 ± 6 | N/A |
| MOF 2 | 110 ± 1 | 72 ± 6 | N/A |

**Table 1**. [a]As-synthesized poly MOFs prior to water treatment. [b]poly MOFs after room temperature water treatment[79].

**2.3 Functionalized MOFs**

In addition to improving the strength of the metal-ligand interaction, MOFs might be functionalized to create steric hindrance to maintain robustness in an aqueous media. By providing hydrophobic pore surfaces or inhibiting the metal ions, it is possible to prevent water molecules from accessing the lattice and damaging the framework structure.

Creating a hydrophobic surface on MOFs by post-synthetic alteration is an effective solution for the water sensitivity of MOFs. The process involves encapsulating hydrophobic guest molecules (e.g., fluorinated compounds[39], polyoxometalates[89], and carbon nanotubes in MOFs pores[90]) and functionalizing their exterior surfaces by ligand replacement and exchange[91] or with carbon coatings by surface thermolysis[92], *etc*[93].

Yang *et al.*[92] discovered a basic technique for considerably enhancing MOFs' moisture resistance. The heat treatment of MOFs resulted in the creation of an amorphous carbon layer on the surface of the MOFs that inhibits hydrolysis. Figure 6A depicts a simple heat treatment at a certain temperature that resulted in coating the frameworks with an amorphous carbon layer, which acted as a moisture repellent when the MOFs were exposed to water, hence inhibiting hydrolysis. Figure 6B(a) demonstrates that the desolvated MOFs were exposed to ambient air (34% relative humidity) and then analyzed by measuring their PXRD patterns to acquire insight into the crystal structure in the presence of moisture. 14 days of exposure to desolvated IRMOF-1 to air (34% relative humidity) led to a shift in relative peak intensities and the emergence of new peaks, suggesting the initiation of hydrolysis.

In contrast, the patterns of all thermally modified MOFs remained essentially unchanged, even after 14 days of humidity exposure. The thermally-treated MOFs displayed significantly better $N_2$ adsorption retention after 14 days of exposure to ambient air, demonstrating enhanced stability in the presence of moisture. In contrast, despite being exposed to humidity for 14 days, all thermally altered MOFs' patterns remained mostly unaltered. The thermally-treated MOFs showed noticeably improved $N_2$ adsorption retention, indicating improved stability in the presence of moisture.

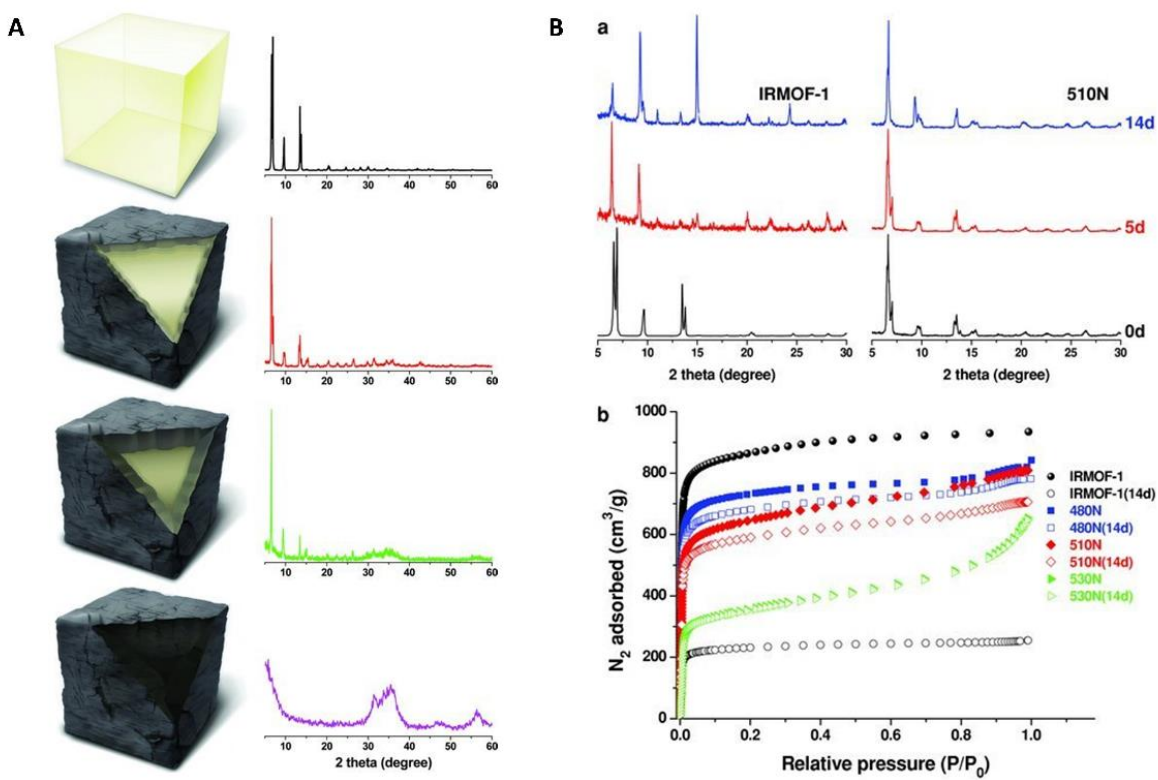

Figure 6 **A.** a Schematic illustration of IRMOF-1 (above) and IRMOF-1 following thermal alteration to create amorphous carbon-coated MOFs (middle) and, at higher temperature, ZnO nanoparticles@amorphous carbon (bottom). The corresponding XRD patterns are shown on the right. **B.** a) PXRD patterns and b) nitrogen adsorption isotherms of IRMOF-1 and the thermally-modified MOFs after exposure to ambient conditions. 480N, 510N, and 530N denote a sample heated at 480 °C, 510 °C, and 530 °C under a nitrogen atmosphere[92].

Numerous studies on the increased hydrothermal stability of MOFs have been published. Taylor *et al.* [94] created CALF-25, a novel porous MOFs. The MOFs has a three-dimensional structure with one-dimensional rectangular pores lined with ethyl ester groups from the ligand. The presence of ethyl ester groups caused the pores to become hydrophobic. The ethyl ester groups inside the pores also prevent CALF-25 against breakdown by water vapor, preserving crystallinity and porosity after exposure to extreme humid environments (90% relative humidity at 353 K).

Omary and his colleagues created a variety of superhydrophobic and water-stable fluorinated metal-organic frameworks (FMOFs)[95-96]. Post-synthetic techniques, such as ligand modification[97] (Figure 7A), and ligand[91, 98] and metal exchange reactions[99] (Figures 7B and 7C), were developed to significantly improve the hydrophobicity and hydrothermal stability of the existing MOFs structures.

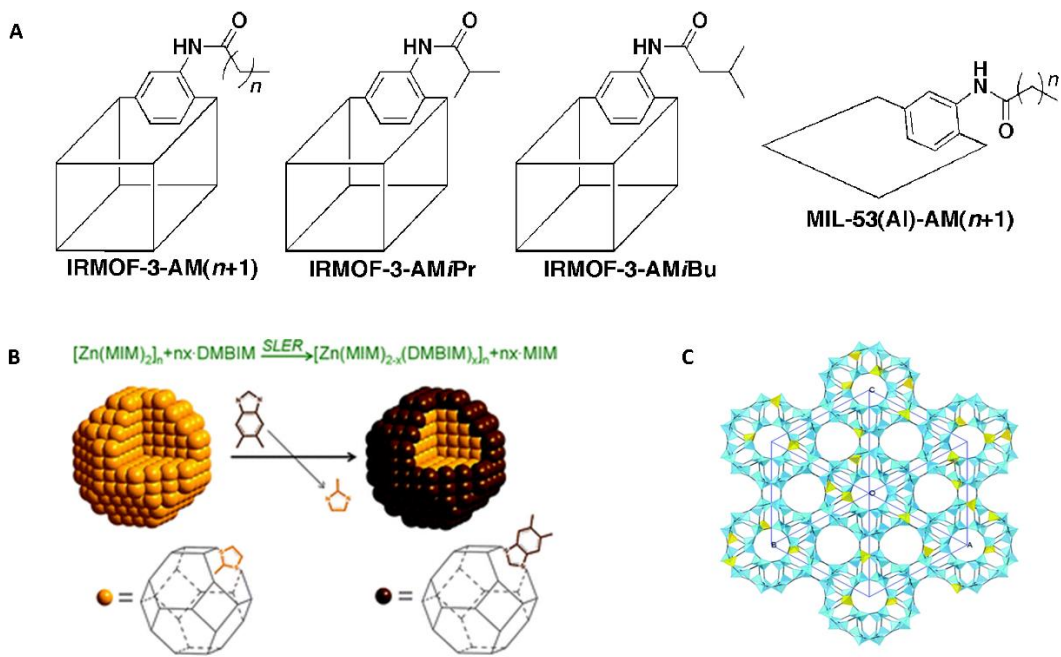

Figure 7. A. Schematic representations of substituted organic ligands[97]. B. Schematic illustration of ZIF-8's shell-ligand-exchange reaction (SLER) mechanism[87]. C. Cyan tetrahedrons and yellow tetrahedrons representing $Zn^{2+}$ and $Cu^{2+}$ nodes, respectively, to demonstrate the doped structure of STU-1[99].

## 3. Coating and supporting of MOFs using hydrophobic materials

The introduction of hydrophobic groups, such as methyls, long alkyl chains, and carbon coatings, would effectively obstruct the pores of MOFs, resulting in a significant reduction in specific surface area and porosity. It is crucial to preserve or even enhance the pore structure of modified MOFs, since it determines the adsorption and diffusion characteristics of adsorbate in porous materials. Coating the crystals of MOFs with hydrophobic polymers to form MOF@polymer composites that display greater structural stability in water-containing conditions than the equivalent MOFs alone is another intriguing post-synthetic modification[100-105]. Methods for stabilization of MOFs in water by hydrophobic materials are shown in Figure 8.

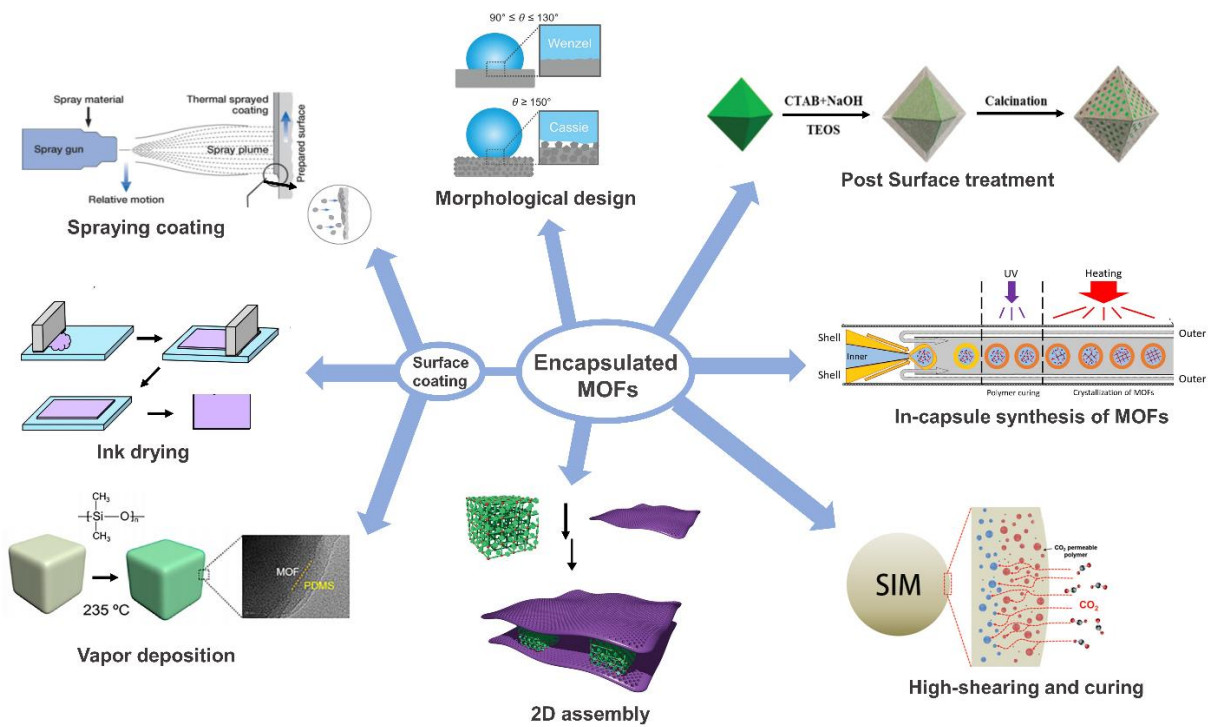

Figure 8. Methods of MOF encapsulation.

Fernandez *et al*. covered the outside surface of MOFs with a layer of hydrophobic Pluronic P123 polymer by physical adsorption[102]. The water stability of polymer-functionalized MOFs is greater than that of unmodified MOFs. However, the contact angle of the coated MOFs is between 22 and 27 degrees, which is insufficient for hydrolysis protection. Jiang *et al*. recently established a comprehensive coating technique to deposit hydrophobic polydimethsiloxane (PDMS) on the surfaces of MOFs, which significantly improved their moisture/water stability[101]. Sun *et al.* also proposed a technique for imparting amphiphobicity to single-crystalline MOFs using 1H,1H,2H,2H-perfluorodecanethiol[104]. Qian *et al*. proposed a solution-immersion procedure for depositing a layer of hydrophobic organosilicone on the exterior surface of MOFs with better water stability[106]. Utilizing UV-curable coating technique, Nanoparticle Organic Hybrid Materials (NOHMs)[107-108] have been encapsulated. The technique is applicable to MOFs[109] as well. Because they have the potential to be utilized in the field of $CO_2$ capture, the examples provided below are described in details.

Zhang *et al*. created a broad method to change hydrophobic PDMS on the surface of MOFs materials in order to considerably improve their resistance to moisture or water using a simple vapor deposition technique[101]. Their research demonstrated that the PDMS protective covering had no effect on the existing structure, porosity, or adsorption site use of MOFs. This unique coating method might make MOFs useful in the presence of water or humidity in a variety of applications, including gas sorption and catalysis, where they have not been utilized previously. MOF-5, HKUST-1, and ZnBT, which are archetypal susceptible MOFs, were effectively coated with PDMS, and the coated samples retained their original crystalline nature and pore properties. Nearly 100% percent of the surface regions of these MOFs were preserved after PDMS coating.

Typically, the coating procedure involves heating MOFs in the presence of PDMS stamps in a sealed glass container at 235 °C, which is appropriate for the majority of MOFs (which are typically stable up to 300 °C). The volatile and low-molecular-weight silicone molecules, which emerge from the thermal decomposition of PDMS, would settle on the surface of MOFs and then cross-link to produce a hydrophobic silicone coating. The coated MOFs would be produced upon cooling to room temperature.

A high-resolution transmission electron microscopy (HRTEM) picture reveals a 10 nm PDMS coating layer on the surface of the MOFs in Figure 9A. It is well knowledge that MOF-5, HKUST-1, and ZnBT are all water-sensitive to variable degrees and are hydrophilic materials with water contact angles near 0°. All PDMS-coated samples have water contact angles of $130 \pm 2°$, indicating their hydrophobic nature (Figure 9B, a–f). Figure 9C depicts the $N_2$ sorption isotherms for three kinds of uncoated and PDMS-coated MOFs before to and after treatment with moisture/water. After exposure to water, the amount of $N_2$ adsorbed by uncoated MOFs has reduced dramatically, but the amount adsorbed by PDMS-coated MOFs remains consistent.

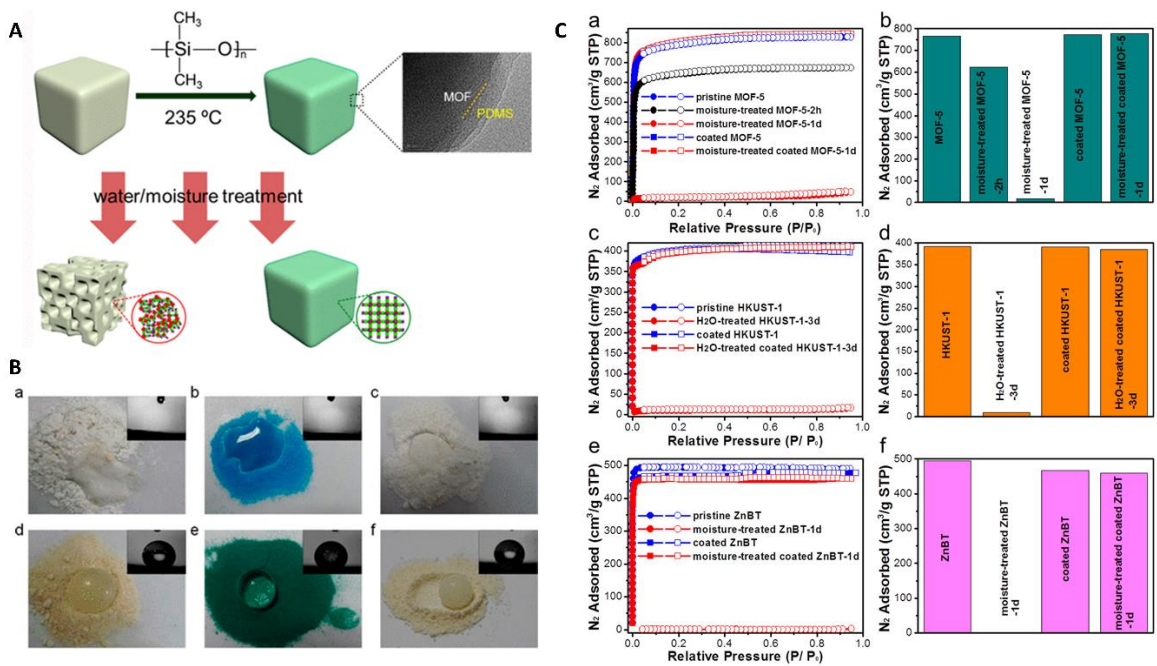

Figure 9. **A.** Illustration of PDMS-Coating on the surface of MOFs and the enhancement of moisture resistance of MOFs. **B.** (a–f) Digital photographs of (a) MOF-5, (d) coated MOF-5, (b) HKUST-1, (e) coated HKUST-1, (c) ZnBT, and (f) coated ZnBT after a drop of water was placed onto the samples. Inset shows contact angle measurement for each sample, respectively (coated MOF-5, 128°; coated HKUST-1, 130°; coated ZnBT, 130°). **C.** (a, c, and e) $N_2$ sorption isotherms for pristine and PDMS-coated MOFs before and after moisture/water treatment (MOF-5 and ZnBT were treated in the air at 55% relative humidity for 1 d; HKUST-1 was treated in water for 3 d). (b, d, and f) $N_2$ sorption capacity of samples at relatively low pressure ($P/P_0$ equal to 0.1)[101].

Qian *et al.* established a simple solution-immersion method for depositing a hydrophobic coating on the outside surface of MOFs without obstructing the MOFs' inherent pores[106]. The method can be performed at room temperature and without heating. Three exemplary MOFs, $NH_2$-MIL-125(Ti), ZIF-67, and HKUST-1, have been chosen for in-depth examination to highlight the adaptability of the method. Figure 10A (a)-(c) demonstrate that water droplets may spread and absorb rapidly on MOFs. (d)-(f) demonstrates that spherical water droplets may stand freely on hydrophobic coating samples without being absorbed. Figure 10B (a), (d), and (g) depict the morphology of three MOF types before to exposure to water. After five days of exposure to water, the crystals in (b), (e), and (h) are almost ruined and full of fractures and voids. (c), (f), and (i) demonstrate that the hydrophobic coating on the surface of MOFs is well retained, as is their crystal shape.

Figure 10C depicts a type I isotherm for the $N_2$ sorption and desorption isotherms of three as-synthesized (AS) MOFs at 77 K. Five days after being exposed to water, the $N_2$ adsorption and desorption curves of three AS MOFs had diminished. $N_2$ sorption curves for hydrophobic-coated MOFs (SS) indicate just a modest reduction. Approximately 85% of capacity remains after the same time span. Figure 10D depicts the similar pattern for the adsorption and desorption curves of $CO_2$.

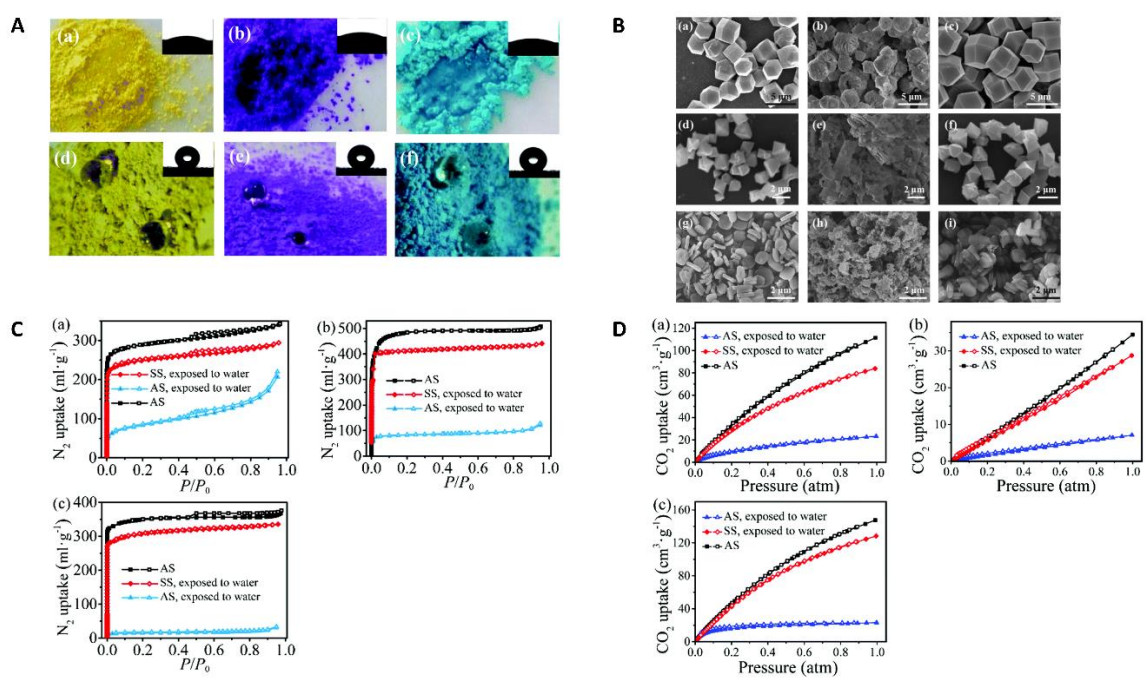

Figure 10. **A.** Digital photographs of the as-synthesized samples (a) $NH_2$-MIL-125(Ti), (b) ZIF-67, (c) HKUST-1, and the SH (d) $NH_2$-MIL-125(Ti), (e) ZIF-67, (f) HKUST-1. The inset shows the contact angle image. **B.** SEM images of (a) the AS ZIF-67, (b) and (c) the AS ZIF-67 and SH ZIF-67 after exposure to water for 5 days, respectively, (d) the AS HKUST-1, (e) and (f) the AS HKUST-1 and SH HKUST-1 after exposure to water for 5 days, respectively, (g) the AS NH2-MIL-125(Ti), (h) and (i) the AS NH2-MIL-125(Ti) and SH NH2-MIL-125(Ti) after exposure to water for 5 days, respectively. **C.** $N_2$ sorption isotherms for (a) NH2-MIL-125(Ti), (b) ZIF-67 and (c) HKUST-1 after exposure to liquid water for 5 days. **D.** $CO_2$ sorption isotherms for (a) $NH_2$-MIL-125(Ti), (b) ZIF-67 and (c) HKUST-1 after exposure to liquid water for 5 days[106].

As illustrated in Figure 11A, Carne-Sanchez et al.[100] demonstrated a technique of spray-dry encapsulation of HKUST-1 crystals in polystyrene microspheres to produce water-resistant composites that preserve the majority of HKUST-1's superior gas sorption capability. The MOFs crystals encased in a polymeric matrix do not require any purifying or filtering operations because the composites are formed in a dry, pure state (Figure 11B). The $N_2$ isotherm at 77 K demonstrated that the composite's value after incubation in water was around 80 percent (Figure 11C). In contrast, HKUST-1 crystals entirely lost their sorption ability under equal circumstances.

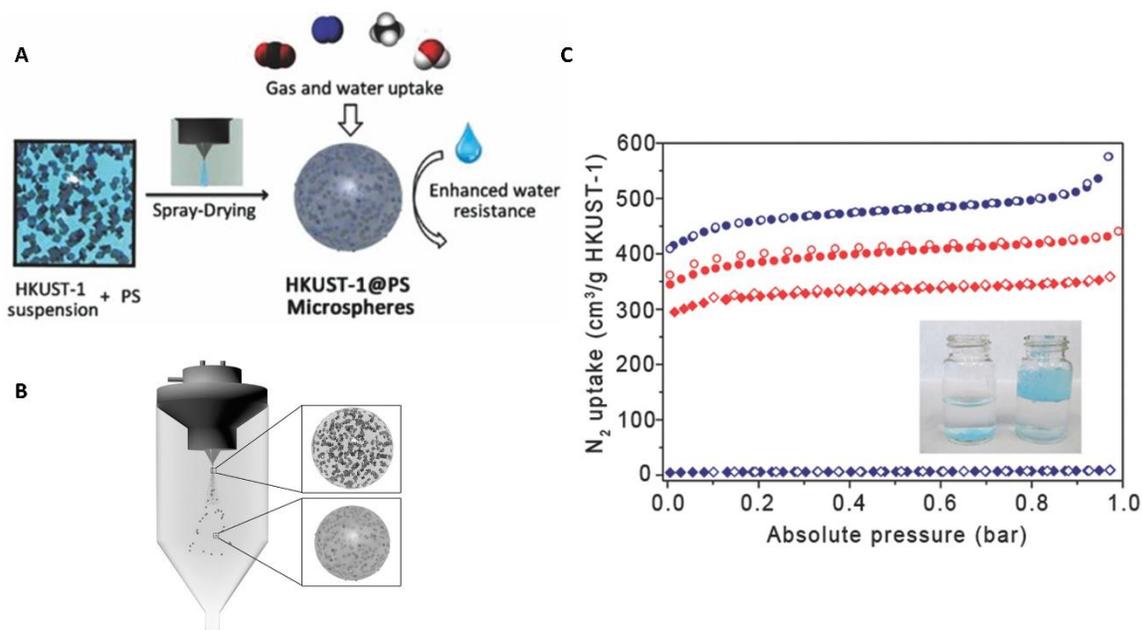

Figure 11. **A.** Spray-dry encapsulation of HKUST-1 crystals into polystyrene microspheres. **B.** Schematic of the spray-drying synthesis of HKUST-1@PS composites. **C.** $N_2$ isotherms at 77 K of HKUST-1 (blue) and HKUST-1@PS_63 (red) before (circles) and after (diamonds) incubation in water[100].

## 4. Water-stable MOFs and Encapsulated MOFs for CO$_2$ Capture from Wet Flue Gas

Due to the intense interest in employing MOFs as sorbents for lowering greenhouse gas emissions, extensive research has been conducted on $CO_2$ capture. $CO_2$ capture in the presence of water is difficult. Although MOFs are capable of capturing large quantities of $CO_2$ under dry conditions, the capture efficiency is drastically reduced in the presence of water because water molecules compete with $CO_2$ molecules for the same binding sites, and in some cases they can even destroy the MOFs framework[110-111]. For example, after exposure to water vapor, the $CO_2$ adsorption capacity of [$Mg_2$(dobpdc)] decreases by approximately 50% when tested under dry circumstances[110-111].

Despite the fact that water content is frequently harmful to $CO_2$ extraction when utilizing MOFs materials, there are instances where water has a negligible effect. Fracaroli *et al.*[111] employed IRMOF-74-III-$CH_2NH_2$ to selectively collect $CO_2$ at a relative humidity of 65%. The experimental findings demonstrate that this MOFs is exceptionally effective at absorbing $CO_2$ (3.2 mmol of $CO_2$ per gram at 800 Torr). However, IRMOF's structure is not stable after 14 days of exposure to water[92]. In addition, Zhang *et al.* noted that their created Zn-pbdc-12a(bpe) and Zn-pbdc-12a(bpy) display $CO_2$ uptakes of 2.2 and 1.8 mmol/g, respectively, under the condition of 1 bar of $CO_2$, which is extremely near to the uptake values before water vapor treatment[79].

In addition, McDonald and colleagues emphasized that the mmen-$M_2$(dobpdc) (M = Mg, Mn, Fe, Co, Zn) compounds, referred to as 'phase-change' adsorbents, show extremely desired properties for the effective capture of $CO_2$[112]. Figure 12 demonstrates that the Langmuir-type $CO_2$ adsorption behavior may be maintained quite well after exposure to water at different temperatures.

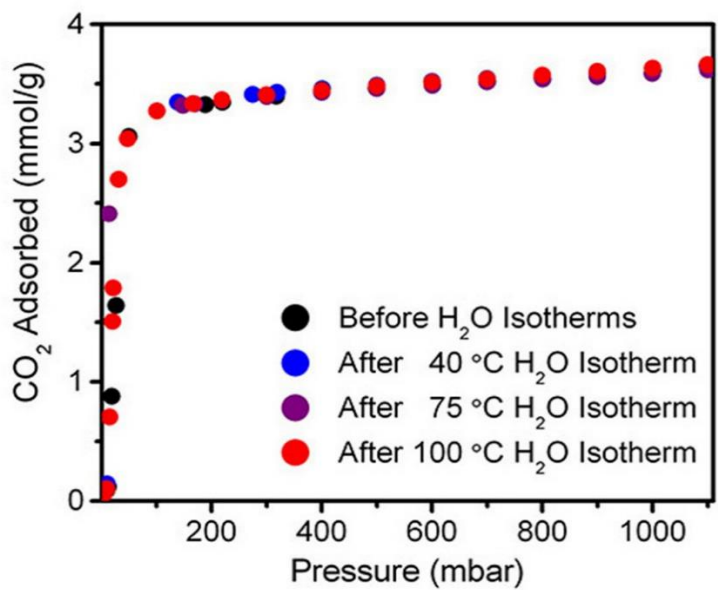

Figure 12. Isothermal adsorption measurements of $CO_2$ with a sample of mmen-$Mg_2$(dobpdc) before exposure to water and after water isotherms at 40, 75 and 100 °C[112].

In addition, Liao et al. [113] shown that functionalizing MOFs with monodentate hydroxide on their pore surfaces may significantly improve $CO_2$ capture performance. The MOFs materials can absorb up to 4.2 mmol/g or 13.4 wt % of $CO_2$ from simulated flue gases, even at high relative humidity (82 %), and then rapidly desorb it under mild regeneration conditions ($N_2$ purge at 358 K). Moon et al. [50] synthesized [$Mg_2$(dobpdc)(DMF)$_2$] and soaked it in polystyrene (PS) solution for one day (150 mg PS in 5 mL N,N-dimethylformamide (DMF)). [$Mg_2$(dobpdc)(DMF)$_2$]@PS was washed three times with DMF to remove excess PS from crystals before being dried under vacuum. One day was spent activating the material under vacuum at 250 °C. Figure 13 depicts the composite's remarkable moisture stability after one day of exposure to 90% humidity.

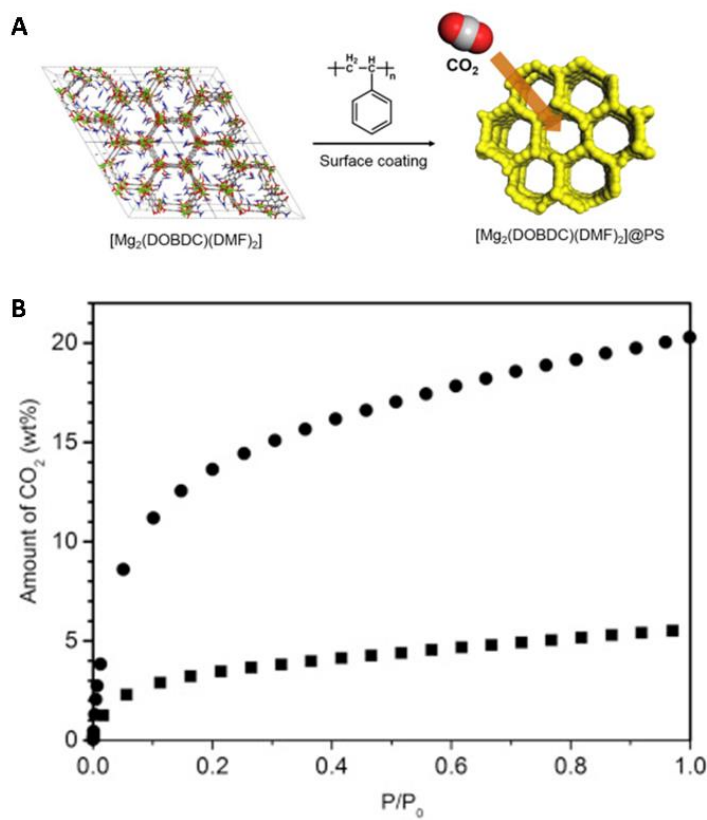

Figure 13. **A**. The structure of [$Mg_2$(dobpdc)(DMF)$_2$]. **B**. $CO_2$ sorption isotherms of [$Mg_2$(dobpdc)(DMF)$_2$] (squares) and [$Mg_2$(dobpdc)(DMF)$_2$]@PS (circles) after humidity exposure[50].

Using an innovative in-situ microencapsulated synthesis, Yu et al.[114] reported the production of a new MOFs-based hybrid sorbent. Figure 14 depicts the production of double emulsions of MOFs precursor solutions and UV-curable silicone shell fluid using a custom-made double capillary microfluidic setup. Microscopic photos of encapsulated HKUST-1 are shown in Figure 15. As illustrated in Figure 15, the geometry and size of the microcapsules are quite uniform[105, 114]. Depending on the system parameters of the microfluidic device and the fluid characteristics, the average capsule diameter is between 300 and 500 microns and the shell thickness is between 35 and 75 microns.

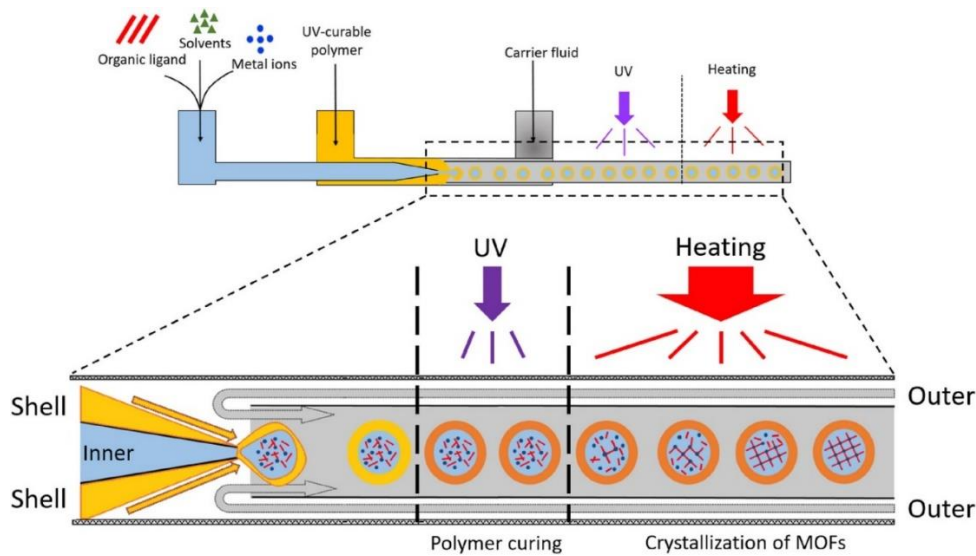

Figure 14. Schematic of the microfluidic system for in-situ encapsulated synthesis of MOFs[114].

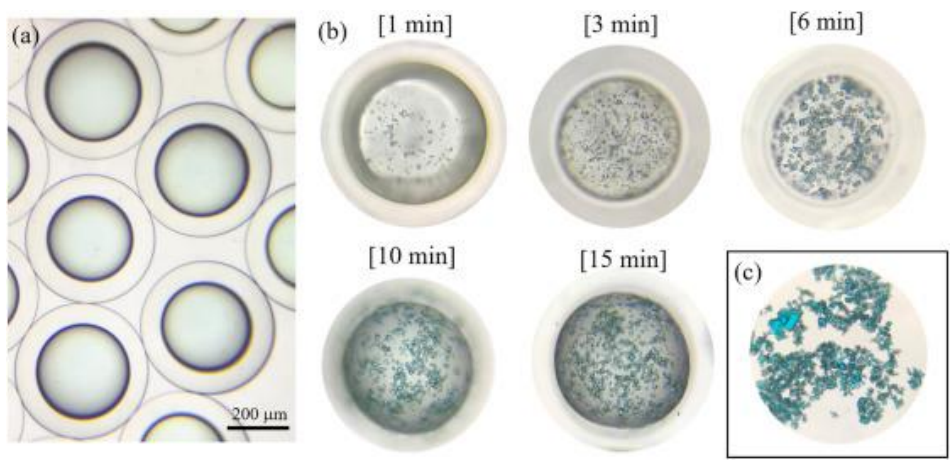

Figure 15. Images of HKUST-1 contained within a capsule. (a) Produced microcapsules containing HKUST-1 precursor solutions; (b) The stages of HKUST-1 crystal formation during the in-situ

thermal reaction within the microcapsules (reaction time: 1–15 min); (c) A comparison image of HKUST-1 crystals synthesized via the conventional solvothermal method using a bulk solution[114].

Consequently, HKUST-1 is effectively produced within the encapsulated droplets of gas-permeable microcapsules. Through TGA analysis, the $CO_2$ capture behavior of encapsulated HKUST-1 is examined. As demonstrated in Figure 16, the $CO_2$ capacity is up to 4 mmol $CO_2$/g MOFs and remains stable after 10 cycles. Microencapsulation is regarded as a potential technique to enhance the stability of sorbents in humid surroundings for $CO_2$ capture[16, 115].

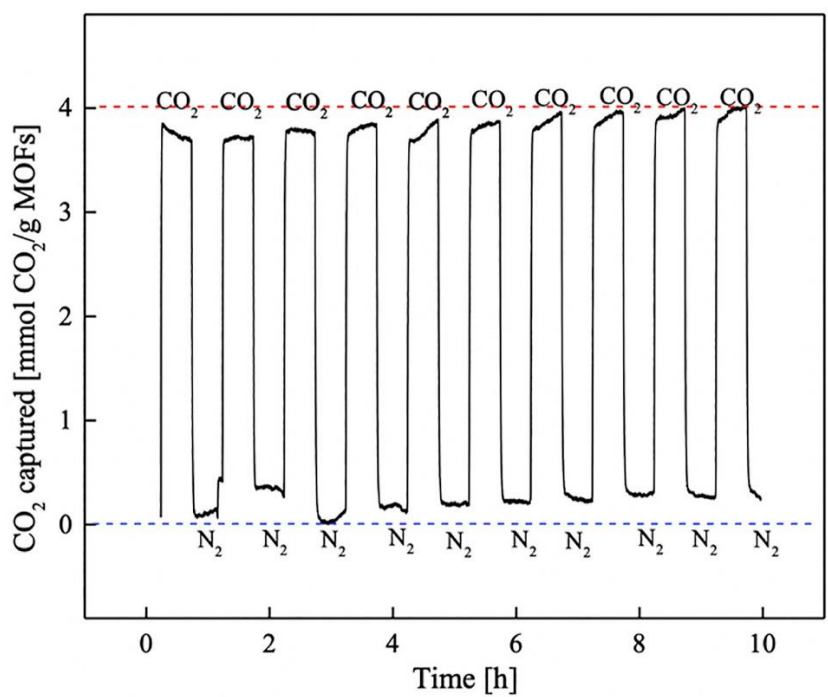

Figure 16. Cyclic study of $CO_2$ capture by encapsulated HKUST-1[114].

## 5. MOFs and Encapsulated MOFs for Direct Air Capture of $CO_2$

According to estimates, only 47% of $CO_2$ emission sources are amenable to traditional carbon capture and storage[116], with the remainder coming from diffuse sources such as isolated industrial sites and transportation. This has led to study interest in the contentious topic of $CO_2$ collection directly from the atmosphere or direct air capture (DAC). Lackner suggested the idea of trapping $CO_2$ from the air for climate change mitigation in 1999[117]. DAC has acquired tremendous momentum during the past several years[118-147]. However, acceptability of these technologies remains restricted[148-149]. Cost and technical difficulties have been cited as important problems. The idea is now widely defined as the direct air capture (DAC) of $CO_2$, and the IPCC created the name "Negative Emission" for such technologies[150]. Several approaches employing various media have been described for DAC, including: physical sorption[151-157], sorption by strong bases[118, 120, 151, 158-164], sorption by amine-modified materials in a thermal-swing process[5, 165-172], sorption by aqueous amino acid solution and followed by precipitation into a guanidine compound[173-181], sorption by ion exchange resin in a moisture-swing process[128, 149, 182-196], and also sorption by polyanthraquinone-carbon nanotube composite in a electrochemical-swing process[197-201].

It is essential to develop sorbent materials for $CO_2$ sorption from ambient air in order to mitigate climate change and complete the carbon cycle. The utilization of sorbent materials, such as MOFs[202-205], is key to most direct air capture methods. However, few research has investigated the use of MOFs for Direct Air Capture of $CO_2$. Unlike $CO_2$ capture from high concentration streams[112, 206-209], $CO_2$ is diluted in the atmosphere. Any procedure for collecting $CO_2$ must avoid using considerable amounts of energy on $CO_2$ removal from bulk air. This excludes heating, cooling, and compressing air. Also, the issue of MOFs' stability in the presence of water must be resolved.

## 5.1 MOFs for Physisorption of $CO_2$:

In the presence of physical sorbents, the interactions between $CO_2$ and the surface of the material are controlled by van der Waals or ion quadrupole forces. Because physisorption relies on physical interactions to bind $CO_2$, this process normally takes place on the surface and inside the pores of a sorbent[152]. This indicates that materials having a large surface area, such as those with a high porosity or nanoscale dimensions[153-155], are desirable. Among these substances are zeolites, activated carbon, alumina, and metal-organic frameworks[151]. However, surface area alone cannot ensure significant $CO_2$ physisorption capacity.

Regeneration with solid sorbents by physisorption requires less energy than regeneration with typical amine solvents. On the other hand, the thermodynamic drivers for $CO_2$ capture are diminished, making physisorption at ambient $CO_2$ levels problematic due to losses in selectivity and absorption capacity. Water vapor is common in ambient air, and MOFs have a very strong affinity for it. Therefore, water's competitive adsorption can greatly lower their $CO_2$ absorption.

As illustrated in Figure 17, Zaworotko and colleagues investigated the $CO_2$ sorption capabilities of zeolite 13X, tetraethylenepentamine-impregnated SBA-15, and microporous and ultramicroporous MOFs physisorbents. Results indicate that physisorbents are capable of capturing $CO_2$ from $CO_2$-rich gas mixtures, shown as Table 2; however, competition and reactivity with ambient moisture greatly limited their DAC performance. The $CO_2$ and $H_2O$ sorption data indicated that controlling the pore size and pore chemistry by crystal engineering may be an effective technique for enhancing $CO_2$ capture performance. The quicker and less energy-intensive recycling of physisorbents might compensate for the lower uptake values, which is a benefit of physisorbents.

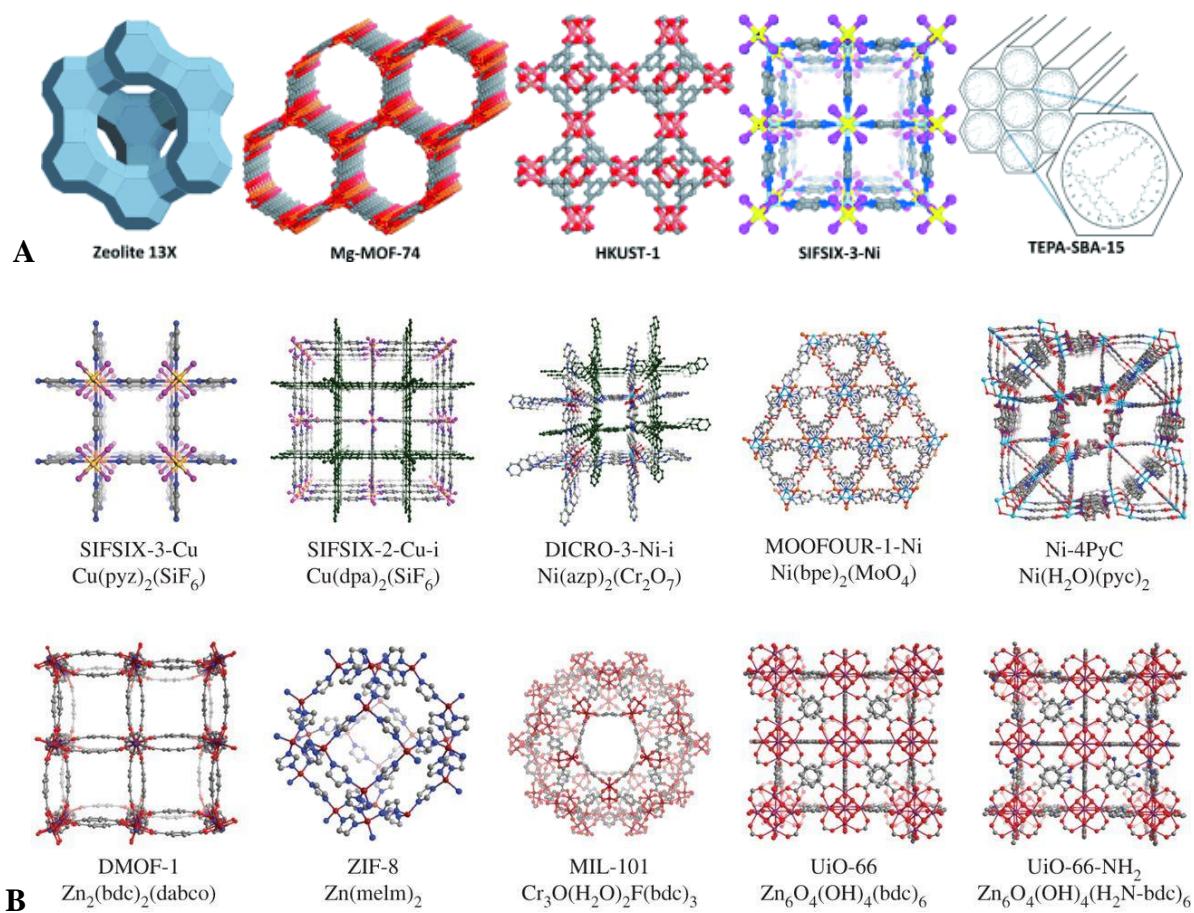

Figure 17. Representative structures of sorbent materials investigated by **A** Amrit Kumar et al.[210] and **B** David G. Madden et al.[211]

| Porous Materials | Capacity (mmol $CO_2$/g) | Ref. |
|---|---|---|
| SIFSIX-3-Ni | 0.182 | 210 |
| HKUST-1 | 0.048 | |
| Mg-MOF-74 | 0.143 | |
| Zeolite 13X | 0.034 | |
| SIFSIX-3-Cu | 0.320 | 211 |
| DICRO-3 | 0.043 | |
| SIFSIX-2-Cu-i | 0.036 | |
| MOOFOUR-1-Ni | 0.056 | |
| Ni-4-PyC | 0.075 | |
| DMOF-1 | 0.030 | |
| ZIF-8 | 0.052 | |
| MIL-101 | 0.023 | |
| UiO-66 | 0.016 | |

**Table 2.** Physisorption Materials under 400 ppm $CO_2$ and 49% relative humidity under 1 atm pressure at room temperature.

Typically, the MOFs with high $CO_2$ temperatures of adsorption featured positively charged open-metal sites or metaloxygen metal bridges. This resulted in powerful electrostatic interactions between the positively charged sites and the negatively charged O atoms in both $CO_2$ and $H_2O$. Findley *et al.*[212] tested the computationally-ready, experimental metal-organic framework (CoRE MOFs) database from 2014 for DAC in humid circumstances, as seen in Figure 18. The performance of zeolite in terms of heat of adsorption is superior to that of MOFs.

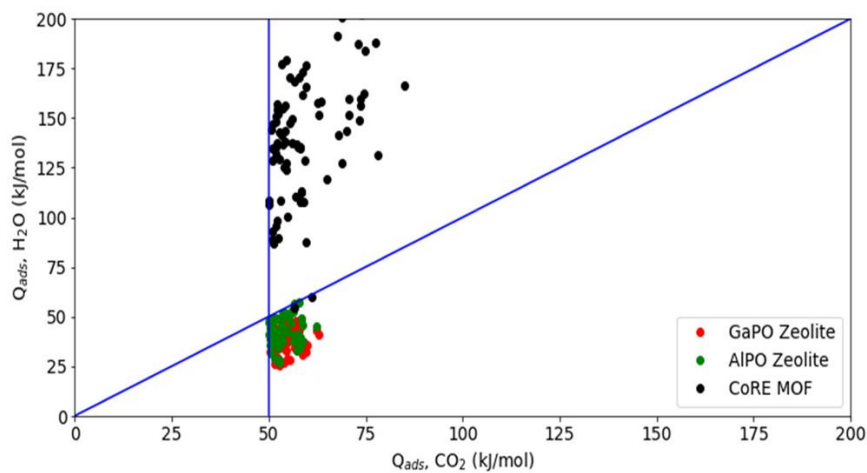

Figure 18. Parity plot showing $CO_2$ and $H_2O$ heats of adsorption at zero loading in GAPO zeolites (red), AlPO zeolites (green), and CoRE MOFs (black). Adsorbents to the right of the vertical line (50 kJ/mol) and below the parity line (blue) were considered further for direct air capture.

Adjusting the pore size of MOFs might maximize physisorption. SIFSIX-3-Cu was tuned to improve the electrostatic van der Waals contact between $CO_2$ and the SIFSIX pillars, making it one of the most promising MOFs for DAC in this class[213]. Using short pyrazine linkages and Cu, the pore size has been reduced to 3.5 relative to structurally comparable precursors. As illustrated in Figure 19, the F-$CO_2$ distance is short and four fluorine atoms point directly into the square channels, allowing for a high charge density. This MOFs has outstanding adsorption characteristics. At 400 ppm $CO_2$, the material had an adsorption capacity of 1.24 mmol/g and exhibited good carbon selectivity over $N_2$ and $CH_4$. However, the influence of humidity on $CO_2$ collection was not mentioned in the study.

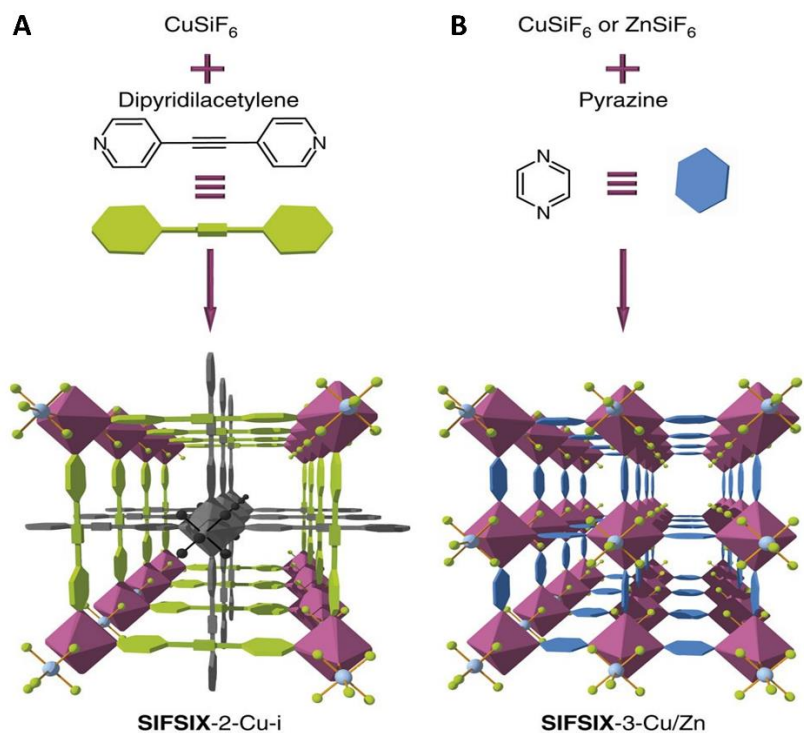

Figure 19. Pore size tuning of the channel structures of **A** SIFSIX-2-Cu-i, **B** SIFSIX-3-Zn or/and SIFSIX-3-Cu. **A** SIFSIX-2-Cu-i; pores size 5.15 Å. **B** SIFSIX-3-Zn; pores size 3.84 Å, BET apparent surface area 250 m$^2$ g$^{-1}$ ; SIFSIX-3-Cu; pores size 3.50 Å[213].

Dynamic breakthrough gas studies conducted on the ultramicroporous material SIFSIX-18-Ni indicate trace $CO_2$ removal from humid air (1000 to 10,000 ppm). Mukherjee *et al.*[214] credit the success of SIFSIX-18-Ni to two normally mutually contradictory factors: a new sort of strong $CO_2$ binding site and hydrophobicity comparable to ZIF-8. SIFSIX-18-Ni also has rapid sorption kinetics to permit selective capture of $CO_2$ over $N_2$ and $H_2O$, making it a prototype for a hitherto unidentified family of physisorbents that display successful trace $CO_2$ capture in both dry and wet environments.

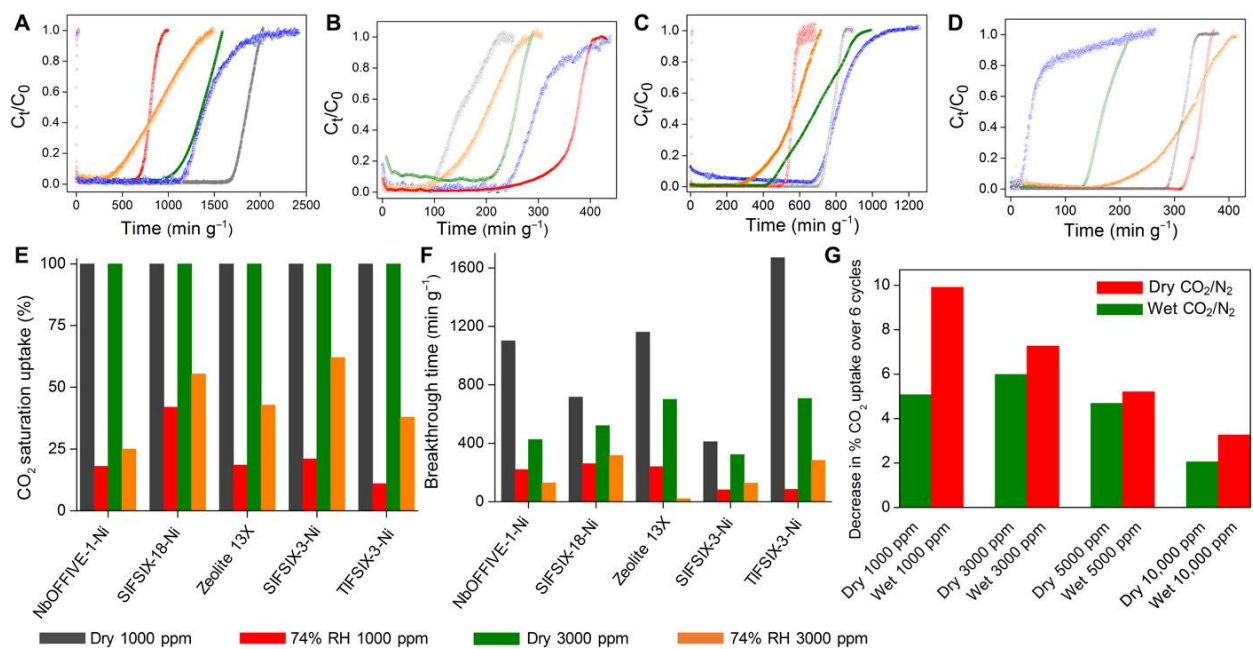

Figure 20. Dynamic gas breakthrough and recyclability tests. Dynamic gas breakthrough tests for SIFSIX-18-Ni-β (red), NbOFFIVE-1-Ni (green), Zeolite 13X (blue), SIFSIX-3-Ni (orange), TIFSIX-3-Ni (gray), and ZIF-8 (purple) using (A) dry 1000 ppm, (B) 74% RH 1000 ppm, (C) dry 3000 ppm, and (D) 74% RH 3000 ppm $CO_2/N_2$ [v/v = 0.1/99.9% for (A) and (B) and v/v = 0.3/99.7% for (C) and (D)] gas mixtures (298 K; 1 bar; flow rate, 20 cm$^3$ min$^{-1}$). (E) Bar diagram exhibiting the relative decline in $CO_2$ saturation uptakes (%) of SIFSIX-18-Ni-β versus other physisorbents (dry/74% RH, 1000/3000 ppm $CO_2/N_2$). (F) Bar diagram of $CO_2$ retention times (min g$^{-1}$) under dry/74% RH, 1000/3000 ppm $CO_2/N_2$. (G) Decrease in % $CO_2$ uptakes over six consecutive adsorption-desorption cycles for SIFSIX-18-Ni-β ($CO_2/N_2$ dry/wet gas mixtures of the following composition: 1000, 3000, 5000, and 10,000 ppm $CO_2$, without/with 74% RH, saturated with $N_2$) [214].

## 5.2 Amine-modified MOFs:

Direct Air Capture of $CO_2$ has been proposed utilizing amines and polyamines on solid supports. Due to their low energy consumption, chemical stability, and great reversibility, amine-modified solid materials have exhibited substantial promise for $CO_2$ extraction from ambient air. There are two distinct sorption mechanisms: (1) Under dry circumstances, the reaction between primary and secondary amines and $CO_2$ produces carbamate or carbamic acid[215-218]. (2) Under circumstances of humidity, amines react with $CO_2$ to form bicarbonate[219-221]. Under dry and wet circumstances, secondary amines $R_1R_2NH$ react with $CO_2$ according to Equations 1 and 2[222].

$$2R_1R_2NH + CO_2 \Leftrightarrow (R_1R_2NH_2^+)(R_1R_2NCOO^-) \Leftrightarrow (R_1R_2NH)(R_1R_2NCOOH) \quad (1)$$

$$R_1R_2NH + CO_2 + H_2O \Leftrightarrow (R_1R_2NH_2^+)(HCO_3^-) \quad (2)$$

The most common methods for the preparation of supported amine sorbents are **(1)** impregnation of amines on porous materials, such as silica, alumina, activated carbon, MOFs, zeolites, and clays[223-242]. This review focuses on MOFs[202-205]. **(2)** The method of functionalization of MOFs for post-combustion $CO_2$ collection and Direct Air Capture has been understood for a very long time[137]. Particular MOFs are synthesized by coordinating certain solvent molecules, often DMF or $H_2O$, to the metal core. These can be eliminated by heating to around 400 degrees Celsius, leaving coordinately unsaturated areas (CUS). In typical MOFs carbon capture, these regions serve as adsorption sites, as the core metal is a strong Lewis acid. These CUS have been used to enable alkylamine-functionalization of DAC. After removing the solvent molecules, the MOFs is put in an organic solvent containing the alkylamine of choice, and the CUS are functionalized to variable degrees depending on the framework. The selected diamines are very basic and firmly coordinate to Lewis acidic metal centers, leaving one amine end unoccupied.

Choi *et al.*[168] introduced pendent amines into the MOFs micropores by modifying MOFs Mg/dobpdc with ethylene diamine (ED). Over four adsorption-desorption cycles with temperature fluctuation, the $CO_2$ capture below 390 ppm was evaluated. ED-Mg-MOF-74 had an adsorption capacity from 1.51 mmol/g to 1.55 mmol/g with no significant change over four cycles, which was higher than non-functionalized Mg-MOF-74 (which also experienced a 20% decrease in adsorption capacity) and close to the final value of PEI-silica at 1.65 mmol/g, which experienced a 29% decrease from 2.36 mmol/g.

Lee *et al.*[110] shown that an ethylenediamine-functionalized metal-organic framework (MOFs) can collect $CO_2$ from ambient air. As illustrated in Figure 21, the MOFs is mmen-Mg$_2$-(dobpdc), which is an equivalent of Mg-MOF-74 with expanded linker ligands (4,4'-dioxido-3,3'-biphenyldicarboxylate) that increased the pore space from 11 to 18.4. The $CO_2$ absorption capability of the sorbent is 2.83 mmol/g from a 390 ppm $CO_2/N_2$ combination at 25 °C. This study also revealed that amine-functionalized MOFs are not significantly affected by humidity, with en-Mg₂(dobpdc) and mmen-Mg₂(dobpdc) exhibiting near-perfect performance after capturing dilute CO₂ from humid air, whereas Mg-MOF-74 experienced a significant decrease, likely due to water molecules binding to CUS and being difficult to remove. However, temperature swing research revealed a 6% decrease in adsorption capacity following 5 cycles.

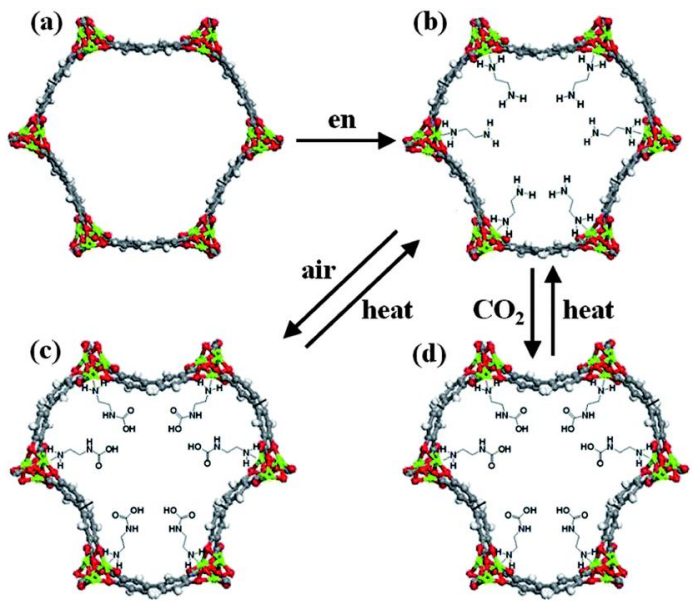

Figure 21. Structures of (a) activated Mg₂(dobpdc) (M=Zn,Mg; dobpdc=4,4′-dioxido-3,3′-biphenyldicarboxylate), (b) en-functionalized Mg₂(dobpdc) (en=ethylenediamine), (c) CO₂ capture from ambient air, and (d) CO₂ capture from flue gas. A reversible transformation between en-functionalized Mg₂(dobpdc) and en-functionalized Mg₂(dobpdc) occurs during heat treatment with CO₂[110].

Darunte *et al.*[243] investigated the kinetics of CO₂ sorption by the sorbent using step-function isotherms. The stepped isotherm is adjustable, and the system may be optimal for DAC applications. CO₂ adsorption in Mg₂(dobpdc) was investigated under ultra-dilute circumstances utilizing a breakthrough adsorption setup as a surrogate for real flow systems. The shape of the isotherm and kinetic parameters significantly lowered the CO₂ capture percentage at low concentrations, hence diminishing the feasibility of Mg₂(dobpdc) for practical DAC applications. Liao *et al.*[244] observed that the CO₂ sorbent reported by Lee *et al.*[110] did not attain saturation at 400 ppm CO₂ in air. This is attributed to the strong intermolecular hydrogen bonds between two neighboring amine groups, which must be broken by higher pressure CO₂[112, 245]. Liao *et al.*[244]

resolved this issue by replacing the ethylenediamine molecule with an even shorter diamine: hydrazine ($H_2N_4$) The authors demonstrate that the novel material [$Mg_2(dobpdc)(H_2N_4)_{1.8}$] absorbs 3.89 mmol/g $CO_2$ at 400 ppm, 298 K, under dry conditions.

**5.3 Potential of water-stable and encapsulated MOFs as moisture-swing $CO_2$ capture sorbent:**

Lackner[128] presented a method for capturing $CO_2$ from ambient air utilizing moisture-swing sorbents in order to offset the high energy cost associated with amine-based sorbents. They bind $CO_2$ in dry environments and release it in moist environments. Water is crucial to the $CO_2$ sorption-desorption mechanism. These sorbents are composed of an anion-exchange resin with quaternary ammonium cations coupled to a polymer network with hydroxide or carbonate anions as mobile counterions. As shown in Figure 22 for a wet resin without $CO_2$ loading, termed State 1 by Shi *et al.*[149], stable carbonate anions are the predominant counterions. As the resin dries, its water content decreases, and the carbonate ion becomes less stable. It causes one of the remaining water molecules to separate into an $HCO_3^-$ ion and an $OH^-$ ion. State 2, which contains OH- ions, has a significant affinity for $CO_2$. The resin absorbs $CO_2$ from air even at low partial pressure. Consequently, State 3 is a completely bicarbonated state. State 4 results from the complete hydration of the bicarbonate ions caused by wetting the resin. This causes $CO_2$ to escape from the moist condition (desorption), which leads to State 1 and the completion of the cycle.

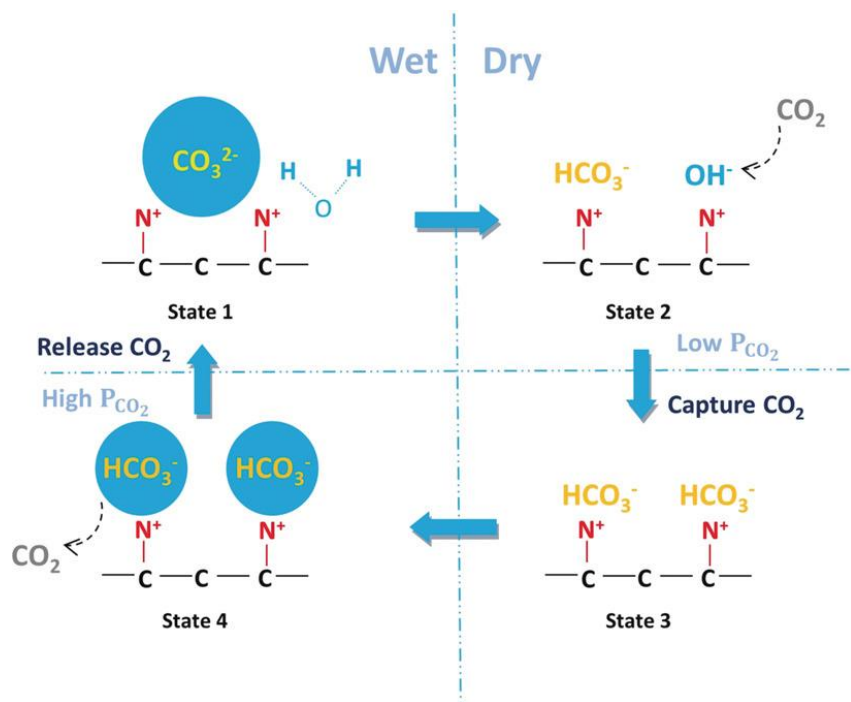

Figure 22. Moisture-swing sorbent for $CO_2$ capture from ambient air[149].

Recently, Shi *et al.* [192] investigated the impacts of sorbent factors on $CO_2$ capture efficiency, paving the door for the future development of sorbents for DAC. Due to their large surface area, porosity, and controllable pore size, water-stable or encapsulated MOFs offer a great deal of promise for use as moisture-swing sorbents. Changing the pore size of MOFs might increase the $CO_2$ absorption capacity of moisture-swing in humid climates.

The $CO_2$ capacity of different types of MOFs discussed in the review under different conditions has been summarized in Figure 23.

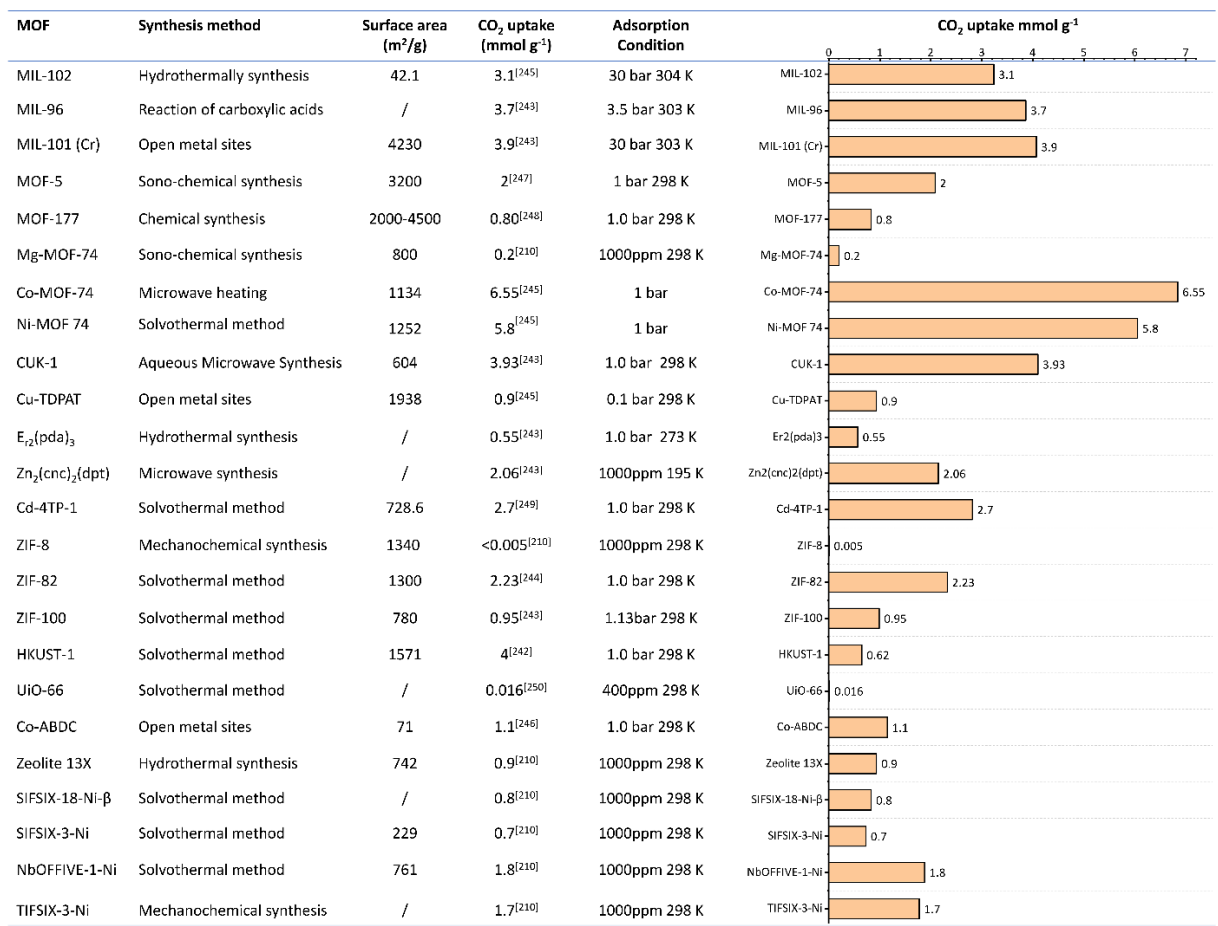

Figure 23: $CO_2$ capacity of different water-stable MOFs.

## 6. Summary and Outlook:

The benefits of MOFs include their form, pore structure, pore size, and surface characteristics adaptability. Temperature and humidity influence the sorption capacity and selectivity of the majority of sorbents. The performance of MOFs in a high-pressure $CO_2$ stream is exceptional, but their sorption capabilities are low when handling gas combinations. Currently, MOFs are badly impacted by water and have water stability difficulties, since the materials experience irreversible structural deterioration in an environment containing water. Therefore, synthesis of water-stable MOFs or encapsulation of MOFs with hydrophobic coatings are essential for $CO_2$ capture by MOFs from wet flue gas or ambient air.

With ongoing attempts to reveal the link between MOF structure and hydro-stability and to investigate methodologies for synthesizing water-stable MOFs, an increasing number of water-stable MOFs have been synthesized. They may be categorized into three groups: metal carboxylate frameworks containing high-valence metal ions, metal azolate frameworks, and MOFs with specialized functionalization. MOFs can also be protected by the hydrophobic coating approach. Certain constraints, such as high-temperature heating, lengthy manufacture, or costly hydrophobic polymers, must be avoided. All MOFs should be compared to one another and to other materials to determine which is now the most cost-effective and which areas require more investigation.

Thus far, technological advancements have been remarkable. However, considerable further study is required. In $CO_2$ capture from wet flue gas, the fundamental demand is for more chemically and thermally resistant materials that can tolerate the high quantities of water in the flue gas stream while simultaneously adjusting the temperature necessary for regeneration. In addition, although the majority of adsorption studies to date have evaluated the performance of materials using single-component $CO_2$ and $N_2$ adsorption isotherms or breakthrough experiments employing a $CO_2/N_2$ mixed gas, a greater understanding of the impact of the presence of water and other minor components in the flue gas is urgently required. For a comprehensive evaluation of the performance of a specific material, it is necessary to examine metal-organic frameworks in parameters that mimic real working conditions (gas composition, temperature, and pressure).

The development of MOFs for Direct Air Capture of $CO_2$ is still in its infancy, with several obstacles left. For the physisorption technique, it is vital to fine tune the MOFs ligands or MOFs metal centers in order to modify their pore size to match the diameter of the $CO_2$ molecule. $CO_2$ content in the air is extremely low. The optimal pore size of MOFs can improve its $CO_2$ storage ability. For the chemisorption process, there are particular enhancements that must be made: the high regeneration energy and synthesis cost of alkylamine-functionalized MOFs must be addressed, although performance is anticipated to increase when other ligands are tried with various frameworks. This is guided by the suggested mechanism, which predicts that altering the distance between surrounding amines and the core metal may increase performance. Sorption of a sorbent for DAC should be performed under actual ambient air conditions (78 percent $N_2$, 21 percent $O_2$, 1 percent inert gases and water vapors, under different humidity conditions). Almost no research has been conducted on MOFs for Direct Air Capture of $CO_2$ under real ambient air conditions.

In addition, if $CO_2$ capture can be paired with its later conversion, the overall sustainability of carbon capture and conversion systems can be substantially enhanced. The creation of catalytic MOFs is an additional field of ongoing study, and MOFs such as Lanthanide MOFs may be constructed to successfully convert $CO_2$ under aqueous-rich and mixed-gas conditions[65]. The effective mass and energy transfer through these materials should be thoroughly explored. MOFs can be designed as hybrid systems that can host both $CO_2$ collection and conversion.

Future research directions of the development of water-stable and encapsulated MOFs for $CO_2$ capture has been shown in Figure 24. In conclusion, all aspects of MOFs for $CO_2$ collection from wet flue gas and ambient air should be evaluated, including sorbent stability, sorption kinetics, sorption capacity, selectivity, regeneration energy penalty, and cost. Developing water-stable, energy-efficient, and low-cost MOFs for $CO_2$ collection is necessary to combat global climate change.

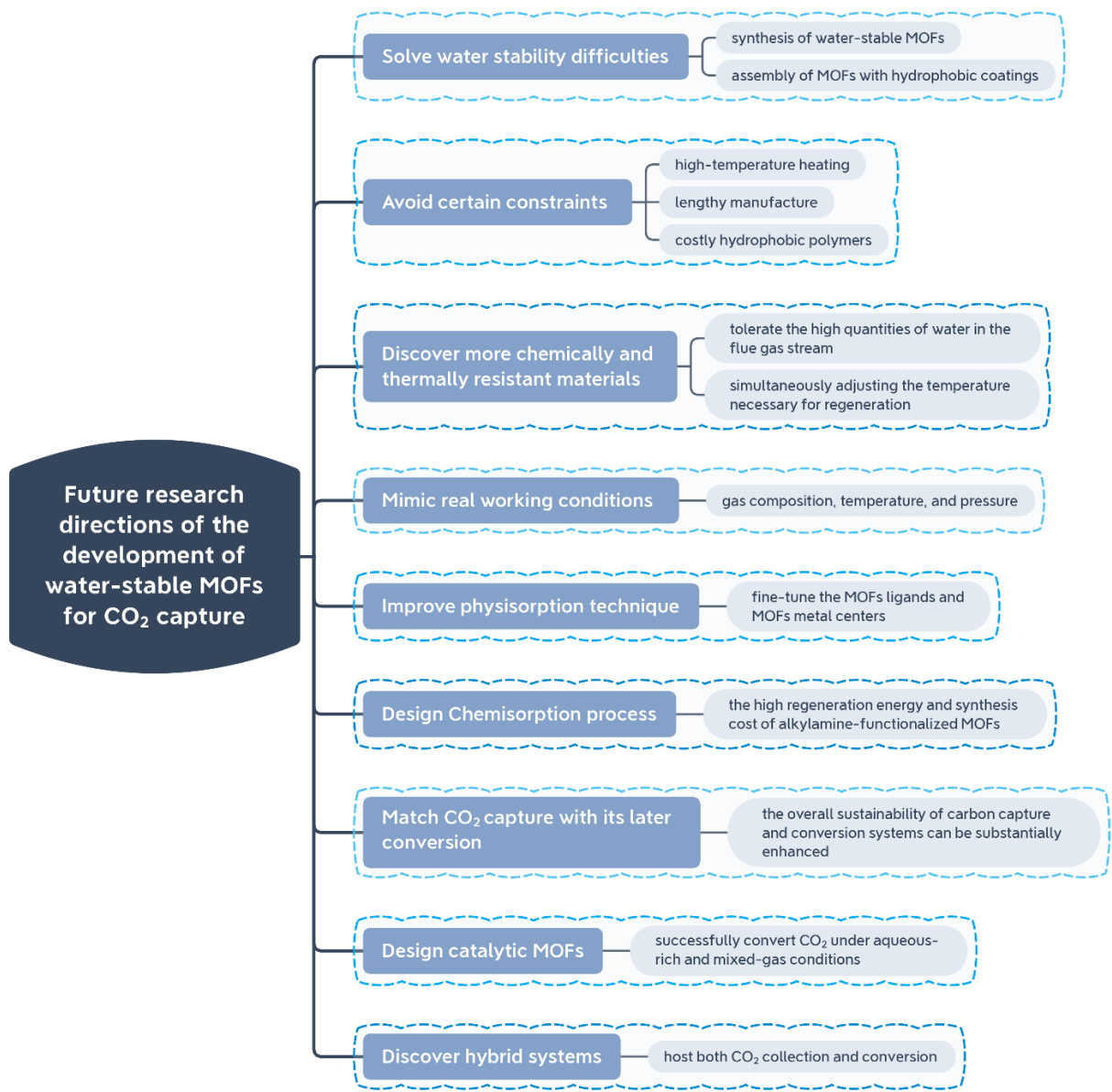

Figure 24. Future research directions of the development of water-stable MOFs for CO$_2$ capture

## Acknowledgment:

L.W. acknowledges the National Key Projects for Research and Development of China (Grant No. 2021YFA1400400), National Natural Science Foundation of China (Grant No. 12074173), and Program for Innovative Talents and Entrepreneur in Jiangsu (Grant No. JSSCTD202101). All authors from Columbia University acknowledge the funding from Saudi Aramco.

## DECLARATION OF INTERESTS

All authors declare no competing interests.

# References:

1. Stocker, T. F.; Qin, D.; Plattner, G.-K.; Tignor, M.; Allen, S. K.; Boschung, J.; Nauels, A.; Xia, Y.; Bex, V.; Midgley, P. M., Climate change 2013: The physical science basis. *Contribution of working group I to the fifth assessment report of the intergovernmental panel on climate change* **2013,** *1535*.
2. Pachauri, R. K.; Allen, M. R.; Barros, V. R.; Broome, J.; Cramer, W.; Christ, R.; Church, J. A.; Clarke, L.; Dahe, Q.; Dasgupta, P., *Climate change 2014: synthesis report. Contribution of Working Groups I, II and III to the fifth assessment report of the Intergovernmental Panel on Climate Change*. IPCC: 2014.
3. Tsouris, C., Separation of $CO_2$ from Flue Gas: A Review AU - Aaron, Douglas. *Sep Sci Technol* **2005,** *40* (1-3), 321-348.
4. Samanta, A.; Zhao, A.; Shimizu, G. K. H.; Sarkar, P.; Gupta, R., Post-Combustion $CO_2$ Capture Using Solid Sorbents: A Review. *Ind. Eng. Chem. Res.* **2012,** *51* (4), 1438-1463.
5. Choi, S.; Drese, J. H.; Jones, C. W., Adsorbent materials for carbon dioxide capture from large anthropogenic point sources. *ChemSusChem* **2009,** *2* (9), 796-854.
6. Flores-Granobles, M.; Saeys, M., Minimizing CO2 emissions with renewable energy: a comparative study of emerging technologies in the steel industry. *Energy Environ. Sci.* **2020,** *13* (7), 1923-1932.
7. Sumida, K.; Rogow, D. L.; Mason, J. A.; McDonald, T. M.; Bloch, E. D.; Herm, Z. R.; Bae, T.-H.; Long, J. R., Carbon Dioxide Capture in Metal–Organic Frameworks. *Chem. Rev.* **2012,** *112* (2), 724-781.
8. Suh, M. P.; Park, H. J.; Prasad, T. K.; Lim, D.-W., Hydrogen Storage in Metal–Organic Frameworks. *Chem. Rev.* **2012,** *112* (2), 782-835.
9. McKinlay, A. C.; Morris, R. E.; Horcajada, P.; Férey, G.; Gref, R.; Couvreur, P.; Serre, C., BioMOFs: Metal–Organic Frameworks for Biological and Medical Applications. *Angew. Chem. Int. Ed.* **2010,** *49* (36), 6260-6266.
10. Thakkar, H.; Eastman, S.; Al-Naddaf, Q.; Rownaghi, A. A.; Rezaei, F., 3D-Printed Metal–Organic Framework Monoliths for Gas Adsorption Processes. *ACS Appl. Mater. Interfaces* **2017,** *9* (41), 35908-35916.
11. Lawson, S.; Rownaghi, A. A.; Rezaei, F., Carbon Hollow Fiber-Supported Metal–Organic Framework Composites for Gas Adsorption. *Energy Technology* **2018,** *6* (4), 694-701.
12. Férey, G.; Serre, C., Large breathing effects in three-dimensional porous hybrid matter: facts, analyses, rules and consequences. *Chem. Soc. Rev.* **2009,** *38* (5), 1380-1399.
13. Lawson, S.; Newport, K.; Pederniera, N.; Rownaghi, A. A.; Rezaei, F., Curcumin Delivery on Metal–Organic Frameworks: The Effect of the Metal Center on Pharmacokinetics within the M-MOF-74 Family. *ACS Applied Bio Materials* **2021,** *4* (4), 3423-3432.
14. Furukawa, H.; Cordova, K. E.; O'Keeffe, M.; Yaghi, O. M., The Chemistry and Applications of Metal-Organic Frameworks. *Science* **2013,** *341* (6149), 1230444.
15. Yaghi, H.-C. Z. J. R. L. O. M., Introduction to Metal–Organic Frameworks. *Chem. Rev.* **2012,** *112* (2), 673-674.
16. Yu, W.; Wang, T.; Park, A.-H. A.; Fang, M., Review of liquid nano-absorbents for enhanced CO2 capture. *Nanoscale* **2019,** *11* (37), 17137-17156.
17. Burtch, N. C.; Jasuja, H.; Walton, K. S., Water Stability and Adsorption in Metal–Organic Frameworks. *Chem. Rev.* **2014,** *114* (20), 10575-10612.

18. Li, T.; Chen, D.-L.; Sullivan, J. E.; Kozlowski, M. T.; Johnson, J. K.; Rosi, N. L., Systematic modulation and enhancement of CO2 : N2 selectivity and water stability in an isoreticular series of bio-MOF-11 analogues. *Chemical Science* **2013,** *4* (4), 1746-1755.
19. Seo, Y.-K.; Yoon, J. W.; Lee, J. S.; Hwang, Y. K.; Jun, C.-H.; Chang, J.-S.; Wuttke, S.; Bazin, P.; Vimont, A.; Daturi, M.; Bourrelly, S.; Llewellyn, P. L.; Horcajada, P.; Serre, C.; Férey, G., Energy-Efficient Dehumidification over Hierachically Porous Metal–Organic Frameworks as Advanced Water Adsorbents. *Advanced Materials* **2012,** *24* (6), 806-810.
20. Férey, G.; Mellot-Draznieks, C.; Serre, C.; Millange, F.; Dutour, J.; Surblé, S.; Margiolaki, I., A Chromium Terephthalate-Based Solid with Unusually Large Pore Volumes and Surface Area. *Science* **2005,** *309* (5743), 2040-2042.
21. Li, Y.; Yang, R. T., Hydrogen storage in metal-organic and covalent-organic frameworks by spillover. *AIChE J.* **2008,** *54* (1), 269-279.
22. Wu, L.; Xiao, J.; Wu, Y.; Xian, S.; Miao, G.; Wang, H.; Li, Z., A Combined Experimental/Computational Study on the Adsorption of Organosulfur Compounds over Metal–Organic Frameworks from Fuels. *Langmuir* **2014,** *30* (4), 1080-1088.
23. Li, H.; Eddaoudi, M.; O'Keeffe, M.; Yaghi, O. M., Design and synthesis of an exceptionally stable and highly porous metal-organic framework. *Nature* **1999,** *402* (6759), 276-279.
24. Kaye, S. S.; Dailly, A.; Yaghi, O. M.; Long, J. R., Impact of Preparation and Handling on the Hydrogen Storage Properties of Zn4O(1,4-benzenedicarboxylate)3 (MOF-5). *J. Am. Chem. Soc.* **2007,** *129* (46), 14176-14177.
25. Cavka, J. H.; Jakobsen, S.; Olsbye, U.; Guillou, N.; Lamberti, C.; Bordiga, S.; Lillerud, K. P., A New Zirconium Inorganic Building Brick Forming Metal Organic Frameworks with Exceptional Stability. *J. Am. Chem. Soc.* **2008,** *130* (42), 13850-13851.
26. Colombo, V.; Galli, S.; Choi, H. J.; Han, G. D.; Maspero, A.; Palmisano, G.; Masciocchi, N.; Long, J. R., High thermal and chemical stability in pyrazolate-bridged metal–organic frameworks with exposed metal sites. *Chemical Science* **2011,** *2* (7), 1311-1319.
27. McHugh, L. N.; McPherson, M. J.; McCormick, L. J.; Morris, S. A.; Wheatley, P. S.; Teat, S. J.; McKay, D.; Dawson, D. M.; Sansome, C. E. F.; Ashbrook, S. E.; Stone, C. A.; Smith, M. W.; Morris, R. E., Hydrolytic stability in hemilabile metal–organic frameworks. *Nat. Chem.* **2018,** *10* (11), 1096-1102.
28. Demir, S.; Bilgin, N.; Cepni, H. M.; Furukawa, H.; Yilmaz, F.; Altintas, C.; Keskin, S., Enhanced water stability and high CO2 storage capacity of a Lewis basic sites-containing zirconium metal–organic framework. *Dalton Transactions* **2021,** *50* (45), 16587-16592.
29. Zhang, L.; Feng, Y.; He, H.; Liu, Y.; Weng, J.; Zhang, P.; Huang, W., Construction of hexanuclear Ce(III) metal–porphyrin frameworks through linker induce strategy for CO2 capture and conversion. *Catalysis Today* **2021,** *374*, 38-43.
30. Zhang, X.; Zhang, K.; Yoo, H.; Lee, Y., Machine Learning-Driven Discovery of Metal–Organic Frameworks for Efficient CO2 Capture in Humid Condition. *ACS Sustainable Chemistry & Engineering* **2021,** *9* (7), 2872-2879.
31. Li, Z.; Wang, L.; Qin, L.; Lai, C.; Wang, Z.; Zhou, M.; Xiao, L.; Liu, S.; Zhang, M., Recent advances in the application of water-stable metal-organic frameworks: Adsorption and photocatalytic reduction of heavy metal in water. *Chemosphere* **2021,** *285*, 131432.
32. Xu, S.; Huang, H.; Guo, X.; Qiao, Z.; Zhong, C., Highly selective gas transport channels in mixed matrix membranes fabricated by using water-stable Cu-BTC. *Separation and Purification Technology* **2021,** *257*, 117979.
33. Guo, M.; Wu, H.; Lv, L.; Meng, H.; Yun, J.; Jin, J.; Mi, J., A Highly Efficient and Stable Composite of Polyacrylate and Metal–Organic Framework Prepared by Interface Engineering for Direct Air Capture. *ACS Appl. Mater. Interfaces* **2021,** *13* (18), 21775-21785.

34. Güçlü, Y.; Erer, H.; Demiral, H.; Altintas, C.; Keskin, S.; Tumanov, N.; Su, B.-L.; Semerci, F., Oxalamide-Functionalized Metal Organic Frameworks for CO2 Adsorption. *ACS Appl. Mater. Interfaces* **2021,** *13* (28), 33188-33198.
35. Peh, S. B.; Farooq, S.; Zhao, D., A metal-organic framework (MOF)-based temperature swing adsorption cycle for postcombustion CO2 capture from wet flue gas. *Chem. Eng. Sci.* **2022,** *250*, 117399.
36. Zhang, S.; Wang, J.; Zhang, Y.; Ma, J.; Huang, L.; Yu, S.; Chen, L.; Song, G.; Qiu, M.; Wang, X., Applications of water-stable metal-organic frameworks in the removal of water pollutants: A review. *Environmental Pollution* **2021,** *291*, 118076.
37. Kaur, H.; Walia, S.; Karmakar, A.; Krishnan, V.; Koner, R. R., Water-stable Zn-based metal-organic framework with hydrophilic-hydrophobic surface for selective adsorption and sensitive detection of oxo-anions and pesticides in aqueous medium. *Journal of Environmental Chemical Engineering* **2022,** *10* (1), 106667.
38. DeCoste, J. B.; Denny, J. M. S.; Peterson, G. W.; Mahle, J. J.; Cohen, S. M., Enhanced aging properties of HKUST-1 in hydrophobic mixed-matrix membranes for ammonia adsorption. *Chemical Science* **2016,** *7* (4), 2711-2716.
39. Decoste, J. B.; Peterson, G. W.; Smith, M. W.; Stone, C. A.; Willis, C. R., Enhanced Stability of Cu-BTC MOF via Perfluorohexane Plasma-Enhanced Chemical Vapor Deposition. *J. Am. Chem. Soc.* **2012,** *134* (3), 1486-1489.
40. Wittmann, T.; Siegel, R.; Reimer, N.; Milius, W.; Stock, N.; Senker, J., Enhancing the Water Stability of Al-MIL-101-NH2 via Postsynthetic Modification. *Chemistry – A European Journal* **2015,** *21* (1), 314-323.
41. Canivet, J.; Fateeva, A.; Guo, Y.; Coasne, B.; Farrusseng, D., Water adsorption in MOFs: fundamentals and applications. *Chem. Soc. Rev.* **2014,** *43* (16), 5594-5617.
42. Howarth, A. J.; Liu, Y.; Li, P.; Li, Z.; Wang, T. C.; Hupp, J. T.; Farha, O. K., Chemical, thermal and mechanical stabilities of metal–organic frameworks. *Nature Reviews Materials* **2016,** *1* (3), 15018.
43. Bon, V.; Senkovskyy, V.; Senkovska, I.; Kaskel, S., Zr(iv) and Hf(iv) based metal–organic frameworks with reo-topology. *Chemical Communications* **2012,** *48* (67), 8407-8409.
44. Bai, Y.; Dou, Y.; Xie, L.-H.; Rutledge, W.; Li, J.-R.; Zhou, H.-C., Zr-based metal–organic frameworks: design, synthesis, structure, and applications. *Chem. Soc. Rev.* **2016,** *45* (8), 2327-2367.
45. Liu, T.-F.; Feng, D.; Chen, Y.-P.; Zou, L.; Bosch, M.; Yuan, S.; Wei, Z.; Fordham, S.; Wang, K.; Zhou, H.-C., Topology-Guided Design and Syntheses of Highly Stable Mesoporous Porphyrinic Zirconium Metal–Organic Frameworks with High Surface Area. *J. Am. Chem. Soc.* **2015,** *137* (1), 413-419.
46. Zhang, M.; Chen, Y.-P.; Bosch, M.; Gentle III, T.; Wang, K.; Feng, D.; Wang, Z. U.; Zhou, H.-C., Symmetry-Guided Synthesis of Highly Porous Metal–Organic Frameworks with Fluorite Topology. *Angew. Chem. Int. Ed.* **2014,** *53* (3), 815-818.
47. Feng, D.; Wang, K.; Su, J.; Liu, T.-F.; Park, J.; Wei, Z.; Bosch, M.; Yakovenko, A.; Zou, X.; Zhou, H.-C., A Highly Stable Zeotype Mesoporous Zirconium Metal–Organic Framework with Ultralarge Pores. *Angew. Chem. Int. Ed.* **2015,** *54* (1), 149-154.
48. Deria, P.; Chung, Y. G.; Snurr, R. Q.; Hupp, J. T.; Farha, O. K., Water stabilization of Zr6-based metal–organic frameworks via solvent-assisted ligand incorporation. *Chemical Science* **2015,** *6* (9), 5172-5176.
49. Deria, P.; Gómez-Gualdrón, D. A.; Bury, W.; Schaef, H. T.; Wang, T. C.; Thallapally, P. K.; Sarjeant, A. A.; Snurr, R. Q.; Hupp, J. T.; Farha, O. K., Ultraporous, Water Stable, and Breathing Zirconium-Based Metal–Organic Frameworks with ftw Topology. *J. Am. Chem. Soc.* **2015,** *137* (40), 13183-13190.
50. Moon, S.-Y.; Liu, Y.; Hupp, J. T.; Farha, O. K., Instantaneous Hydrolysis of Nerve-Agent Simulants with a Six-Connected Zirconium-Based Metal–Organic Framework. *Angew. Chem. Int. Ed.* **2015,** *54* (23), 6795-6799.
51. Cadiau, A.; Lee, J. S.; Damasceno Borges, D.; Fabry, P.; Devic, T.; Wharmby, M. T.; Martineau, C.; Foucher, D.; Taulelle, F.; Jun, C.-H.; Hwang, Y. K.; Stock, N.; De Lange, M. F.; Kapteijn, F.; Gascon, J.; Maurin,

G.; Chang, J.-S.; Serre, C., Design of Hydrophilic Metal Organic Framework Water Adsorbents for Heat Reallocation. *Advanced Materials* **2015,** *27* (32), 4775-4780.
52. Mouchaham, G.; Cooper, L.; Guillou, N.; Martineau, C.; Elkaïm, E.; Bourrelly, S.; Llewellyn, P. L.; Allain, C.; Clavier, G.; Serre, C.; Devic, T., A Robust Infinite Zirconium Phenolate Building Unit to Enhance the Chemical Stability of Zr MOFs. *Angew. Chem. Int. Ed.* **2015,** *54* (45), 13297-13301.
53. Zheng, J.; Wu, M.; Jiang, F.; Su, W.; Hong, M., Stable porphyrin Zr and Hf metal–organic frameworks featuring 2.5 nm cages: high surface areas, SCSC transformations and catalyses. *Chemical Science* **2015,** *6* (6), 3466-3470.
54. Kalidindi, S. B.; Nayak, S.; Briggs, M. E.; Jansat, S.; Katsoulidis, A. P.; Miller, G. J.; Warren, J. E.; Antypov, D.; Corà, F.; Slater, B.; Prestly, M. R.; Martí-Gastaldo, C.; Rosseinsky, M. J., Chemical and Structural Stability of Zirconium-based Metal–Organic Frameworks with Large Three-Dimensional Pores by Linker Engineering. *Angew. Chem. Int. Ed.* **2015,** *54* (1), 221-226.
55. Plessius, R.; Kromhout, R.; Ramos, A. L. D.; Ferbinteanu, M.; Mittelmeijer-Hazeleger, M. C.; Krishna, R.; Rothenberg, G.; Tanase, S., Highly Selective Water Adsorption in a Lanthanum Metal–Organic Framework. *Chemistry – A European Journal* **2014,** *20* (26), 7922-7925.
56. Qin, L.; Lin, L.-X.; Fang, Z.-P.; Yang, S.-P.; Qiu, G.-H.; Chen, J.-X.; Chen, W.-H., A water-stable metal–organic framework of a zwitterionic carboxylate with dysprosium: a sensing platform for Ebolavirus RNA sequences. *Chemical Communications* **2016,** *52* (1), 132-135.
57. Liang, Y.-T.; Yang, G.-P.; Liu, B.; Yan, Y.-T.; Xi, Z.-P.; Wang, Y.-Y., Four super water-stable lanthanide–organic frameworks with active uncoordinated carboxylic and pyridyl groups for selective luminescence sensing of Fe3+. *Dalton Transactions* **2015,** *44* (29), 13325-13330.
58. Dong, X.-Y.; Wang, R.; Wang, J.-Z.; Zang, S.-Q.; Mak, T. C. W., Highly selective Fe3+ sensing and proton conduction in a water-stable sulfonate–carboxylate Tb–organic-framework. *Journal of Materials Chemistry A* **2015,** *3* (2), 641-647.
59. Qin, J.; Ma, B.; Liu, X.-F.; Lu, H.-L.; Dong, X.-Y.; Zang, S.-Q.; Hou, H., Aqueous- and vapor-phase detection of nitroaromatic explosives by a water-stable fluorescent microporous MOF directed by an ionic liquid. *Journal of Materials Chemistry A* **2015,** *3* (24), 12690-12697.
60. Cao, L.-H.; Shi, F.; Zhang, W.-M.; Zang, S.-Q.; Mak, T. C. W., Selective Sensing of Fe3+ and Al3+ Ions and Detection of 2,4,6-Trinitrophenol by a Water-Stable Terbium-Based Metal–Organic Framework. *Chemistry – A European Journal* **2015,** *21* (44), 15705-15712.
61. Yao, S.; Wang, D.; Cao, Y.; Li, G.; Huo, Q.; Liu, Y., Two stable 3D porous metal–organic frameworks with high performance for gas adsorption and separation. *Journal of Materials Chemistry A* **2015,** *3* (32), 16627-16632.
62. Dan, W.; Liu, X.; Deng, M.; Ling, Y.; Chen, Z.; Zhou, Y., A highly stable indium phosphonocarboxylate framework as a multifunctional sensor for Cu2+ and methylviologen ions. *Dalton Transactions* **2015,** *44* (8), 3794-3800.
63. Hao, J.-N.; Yan, B., Recyclable lanthanide-functionalized MOF hybrids to determine hippuric acid in urine as a biological index of toluene exposure. *Chemical Communications* **2015,** *51* (77), 14509-14512.
64. Reimer, N.; Bueken, B.; Leubner, S.; Seidler, C.; Wark, M.; De Vos, D.; Stock, N., Three Series of Sulfo-Functionalized Mixed-Linker CAU-10 Analogues: Sorption Properties, Proton Conductivity, and Catalytic Activity. *Chemistry – A European Journal* **2015,** *21* (35), 12517-12524.
65. Le, D. H.; Loughan, R. P.; Gładysiak, A.; Rampal, N.; Brooks, I. A.; Park, A.-H. A.; Fairen-Jimenez, D.; Stylianou, K. C., Lanthanide metal–organic frameworks for the fixation of CO2 under aqueous-rich and mixed-gas conditions. *Journal of Materials Chemistry A* **2022,** *10* (3), 1442-1450.
66. Zhang, J.-P.; Zhang, Y.-B.; Lin, J.-B.; Chen, X.-M., Metal Azolate Frameworks: From Crystal Engineering to Functional Materials. *Chem. Rev.* **2012,** *112* (2), 1001-1033.

67. Banerjee, R.; Phan, A.; Wang, B.; Knobler, C.; Furukawa, H.; O'Keeffe, M.; Yaghi, O. M., High-Throughput Synthesis of Zeolitic Imidazolate Frameworks and Application to $CO_2$ Capture. *Science* **2008,** *319* (5865), 939-943.
68. Park, K. S.; Ni, Z.; Côté, A. P.; Choi, J. Y.; Huang, R.; Uribe-Romo, F. J.; Chae, H. K.; O'Keeffe, M.; Yaghi, O. M., Exceptional chemical and thermal stability of zeolitic imidazolate frameworks. *Proc. Natl. Acad. Sci.* **2006,** *103* (27), 10186-10191.
69. Huang, X.-C.; Lin, Y.-Y.; Zhang, J.-P.; Chen, X.-M., Ligand-Directed Strategy for Zeolite-Type Metal–Organic Frameworks: Zinc(II) Imidazolates with Unusual Zeolitic Topologies. *Angew. Chem. Int. Ed.* **2006,** *45* (10), 1557-1559.
70. Zhang, J.-P.; Zhu, A.-X.; Lin, R.-B.; Qi, X.-L.; Chen, X.-M., Pore Surface Tailored SOD-Type Metal-Organic Zeolites. *Advanced Materials* **2011,** *23* (10), 1268-1271.
71. Liao, P.-Q.; Zhou, D.-D.; Zhu, A.-X.; Jiang, L.; Lin, R.-B.; Zhang, J.-P.; Chen, X.-M., Strong and Dynamic CO2 Sorption in a Flexible Porous Framework Possessing Guest Chelating Claws. *J. Am. Chem. Soc.* **2012,** *134* (42), 17380-17383.
72. He, C.-T.; Tian, J.-Y.; Liu, S.-Y.; Ouyang, G.; Zhang, J.-P.; Chen, X.-M., A porous coordination framework for highly sensitive and selective solid-phase microextraction of non-polar volatile organic compounds. *Chemical Science* **2013,** *4* (1), 351-356.
73. Liao, P.-Q.; Zhang, W.-X.; Zhang, J.-P.; Chen, X.-M., Efficient purification of ethene by an ethane-trapping metal-organic framework. *Nature Communications* **2015,** *6* (1), 8697.
74. He, C.-T.; Jiang, L.; Ye, Z.-M.; Krishna, R.; Zhong, Z.-S.; Liao, P.-Q.; Xu, J.; Ouyang, G.; Zhang, J.-P.; Chen, X.-M., Exceptional Hydrophobicity of a Large-Pore Metal–Organic Zeolite. *J. Am. Chem. Soc.* **2015,** *137* (22), 7217-7223.
75. Fu, H.-R.; Xu, Z.-X.; Zhang, J., Water-Stable Metal–Organic Frameworks for Fast and High Dichromate Trapping via Single-Crystal-to-Single-Crystal Ion Exchange. *Chem. Mater.* **2015,** *27* (1), 205-210.
76. Zhou, E.-L.; Qin, C.; Wang, X.-L.; Shao, K.-Z.; Su, Z.-M., Steam-Assisted Synthesis of an Extra-Stable Polyoxometalate-Encapsulating Metal Azolate Framework: Applications in Reagent Purification and Proton Conduction. *Chemistry – A European Journal* **2015,** *21* (37), 13058-13064.
77. Meng, X.; Zhong, R.-L.; Song, X.-Z.; Song, S.-Y.; Hao, Z.-M.; Zhu, M.; Zhao, S.-N.; Zhang, H.-J., A stable, pillar-layer metal–organic framework containing uncoordinated carboxyl groups for separation of transition metal ions. *Chemical Communications* **2014,** *50* (48), 6406-6408.
78. Wang, J.; Zhang, J.; Jin, F.; Luo, Y.; Wang, S.; Zhang, Z.; Wu, Y.; Liu, H.; Lu, J. Y.; Fang, M., Synthesis of an exceptional water-stable two-fold interpenetrated Zn(ii)-paddlewheel metal–organic framework. *CrystEngComm* **2015,** *17* (31), 5906-5910.
79. Zhang, Z.; Nguyen, H. T. H.; Miller, S. A.; Ploskonka, A. M.; DeCoste, J. B.; Cohen, S. M., Polymer–Metal–Organic Frameworks (polyMOFs) as Water Tolerant Materials for Selective Carbon Dioxide Separations. *J. Am. Chem. Soc.* **2016,** *138* (3), 920-925.
80. Lin, Y.; Zhang, Q.; Zhao, C.; Li, H.; Kong, C.; Shen, C.; Chen, L., An exceptionally stable functionalized metal–organic framework for lithium storage. *Chemical Communications* **2015,** *51* (4), 697-699.
81. Sanda, S.; Biswas, S.; Konar, S., Study of Proton Conductivity of a 2D Flexible MOF and a 1D Coordination Polymer at Higher Temperature. *Inorg. Chem.* **2015,** *54* (4), 1218-1222.
82. Begum, S.; Horike, S.; Kitagawa, S.; Krautscheid, H., Water stable triazolyl phosphonate MOFs: steep water uptake and facile regeneration. *Dalton Transactions* **2015,** *44* (43), 18727-18730.
83. Wang, J.-H.; Li, M.; Li, D., An Exceptionally Stable and Water-Resistant Metal–Organic Framework with Hydrophobic Nanospaces for Extracting Aromatic Pollutants from Water. *Chemistry – A European Journal* **2014,** *20* (38), 12004-12008.

84. Zhou, E.-L.; Qin, C.; Huang, P.; Wang, X.-L.; Chen, W.-C.; Shao, K.-Z.; Su, Z.-M., A Stable Polyoxometalate-Pillared Metal–Organic Framework for Proton-Conducting and Colorimetric Biosensing. *Chemistry – A European Journal* **2015,** *21* (33), 11894-11898.
85. Wang, K.; Lv, X.-L.; Feng, D.; Li, J.; Chen, S.; Sun, J.; Song, L.; Xie, Y.; Li, J.-R.; Zhou, H.-C., Pyrazolate-Based Porphyrinic Metal–Organic Framework with Extraordinary Base-Resistance. *J. Am. Chem. Soc.* **2016,** *138* (3), 914-919.
86. Song, X.-Z.; Song, S.-Y.; Zhao, S.-N.; Hao, Z.-M.; Zhu, M.; Meng, X.; Wu, L.-L.; Zhang, H.-J., Single-Crystal-to-Single-Crystal Transformation of a Europium(III) Metal–Organic Framework Producing a Multi-responsive Luminescent Sensor. *Advanced Functional Materials* **2014,** *24* (26), 4034-4041.
87. Saccoccia, B.; Bohnsack, A. M.; Waggoner, N. W.; Cho, K. H.; Lee, J. S.; Hong, D.-Y.; Lynch, V. M.; Chang, J.-S.; Humphrey, S. M., Separation of p-Divinylbenzene by Selective Room-Temperature Adsorption Inside Mg-CUK-1 Prepared by Aqueous Microwave Synthesis. *Angew. Chem. Int. Ed.* **2015,** *54* (18), 5394-5398.
88. Wang, D.; Zhang, L.; Li, G.; Huo, Q.; Liu, Y., Luminescent MOF material based on cadmium(ii) and mixed ligands: application for sensing volatile organic solvent molecules. *RSC Advances* **2015,** *5* (23), 18087-18091.
89. Mustafa, D.; Breynaert, E.; Bajpe, S. R.; Martens, J. A.; Kirschhock, C. E. A., Stability improvement of Cu3(BTC)2 metal–organic frameworks under steaming conditions by encapsulation of a Keggin polyoxometalate. *Chemical Communications* **2011,** *47* (28), 8037-8039.
90. Yang, S. J.; Choi, J. Y.; Chae, H. K.; Cho, J. H.; Nahm, K. S.; Park, C. R., Preparation and Enhanced Hydrostability and Hydrogen Storage Capacity of CNT@MOF-5 Hybrid Composite. *Chem. Mater.* **2009,** *21* (9), 1893-1897.
91. Liu, X.; Li, Y.; Ban, Y.; Peng, Y.; Jin, H.; Bux, H.; Xu, L.; Caro, J.; Yang, W., Improvement of hydrothermal stability of zeolitic imidazolate frameworks. *Chemical Communications* **2013,** *49* (80), 9140-9142.
92. Yang, S. J.; Park, C. R., Preparation of Highly Moisture-Resistant Black-Colored Metal Organic Frameworks. *Advanced Materials* **2012,** *24* (29), 4010-4013.
93. McGuire, C. V.; Forgan, R. S., The surface chemistry of metal–organic frameworks. *Chemical Communications* **2015,** *51* (25), 5199-5217.
94. Taylor, J. M.; Vaidhyanathan, R.; Iremonger, S. S.; Shimizu, G. K. H., Enhancing Water Stability of Metal–Organic Frameworks via Phosphonate Monoester Linkers. *J. Am. Chem. Soc.* **2012,** *134* (35), 14338-14340.
95. Nijem, N.; Canepa, P.; Kaipa, U.; Tan, K.; Roodenko, K.; Tekarli, S.; Halbert, J.; Oswald, I. W. H.; Arvapally, R. K.; Yang, C.; Thonhauser, T.; Omary, M. A.; Chabal, Y. J., Water Cluster Confinement and Methane Adsorption in the Hydrophobic Cavities of a Fluorinated Metal–Organic Framework. *J. Am. Chem. Soc.* **2013,** *135* (34), 12615-12626.
96. Yang, C.; Kaipa, U.; Mather, Q. Z.; Wang, X.; Nesterov, V.; Venero, A. F.; Omary, M. A., Fluorous Metal–Organic Frameworks with Superior Adsorption and Hydrophobic Properties toward Oil Spill Cleanup and Hydrocarbon Storage. *J. Am. Chem. Soc.* **2011,** *133* (45), 18094-18097.
97. Nguyen, J. G.; Cohen, S. M., Moisture-Resistant and Superhydrophobic Metal−Organic Frameworks Obtained via Postsynthetic Modification. *J. Am. Chem. Soc.* **2010,** *132* (13), 4560-4561.
98. Wu, T.; Shen, L.; Luebbers, M.; Hu, C.; Chen, Q.; Ni, Z.; Masel, R. I., Enhancing the stability of metal–organic frameworks in humid air by incorporating water repellent functional groups. *Chemical Communications* **2010,** *46* (33), 6120-6122.
99. Zhu, X.-W.; Zhou, X.-P.; Li, D., Exceptionally water stable heterometallic gyroidal MOFs: tuning the porosity and hydrophobicity by doping metal ions. *Chemical Communications* **2016,** *52* (39), 6513-6516.

100. Carné-Sánchez, A.; Stylianou, K. C.; Carbonell, C.; Naderi, M.; Imaz, I.; Maspoch, D., Protecting Metal–Organic Framework Crystals from Hydrolytic Degradation by Spray-Dry Encapsulating Them into Polystyrene Microspheres. *Advanced Materials* **2015,** *27* (5), 869-873.
101. Zhang, W.; Hu, Y.; Ge, J.; Jiang, H.-L.; Yu, S.-H., A Facile and General Coating Approach to Moisture/Water-Resistant Metal–Organic Frameworks with Intact Porosity. *J. Am. Chem. Soc.* **2014,** *136* (49), 16978-16981.
102. Fernandez, C. A.; Nune, S. K.; Annapureddy, H. V.; Dang, L. X.; McGrail, B. P.; Zheng, F.; Polikarpov, E.; King, D. L.; Freeman, C.; Brooks, K. P., Hydrophobic and moisture-stable metal–organic frameworks. *Dalton Transactions* **2015,** *44* (30), 13490-13497.
103. Huang, G.; Yang, Q.; Xu, Q.; Yu, S.-H.; Jiang, H.-L., Polydimethylsiloxane Coating for a Palladium/MOF Composite: Highly Improved Catalytic Performance by Surface Hydrophobization. *Angew. Chem. Int. Ed.* **2016,** *55* (26), 7379-7383.
104. Sun, Q.; He, H.; Gao, W.-Y.; Aguila, B.; Wojtas, L.; Dai, Z.; Li, J.; Chen, Y.-S.; Xiao, F.-S.; Ma, S., Imparting amphiphobicity on single-crystalline porous materials. *Nature Communications* **2016,** *7* (1), 13300.
105. Yu, W.; Wang, T.; Park, A.-H. A.; Fang, M., Toward Sustainable Energy and Materials: CO2 Capture Using Microencapsulated Sorbents. *Ind. Eng. Chem. Res.* **2020,** *59* (21), 9746-9759.
106. Qian, X.; Sun, F.; Sun, J.; Wu, H.; Xiao, F.; Wu, X.; Zhu, G., Imparting surface hydrophobicity to metal–organic frameworks using a facile solution-immersion process to enhance water stability for CO2 capture. *Nanoscale* **2017,** *9* (5), 2003-2008.
107. Moore, T.; Rim, G.; Park, A.-H. A.; Mumford, K. A.; Stevens, G. W.; Webley, P. A., Encapsulation of highly viscous CO2 capture solvents for enhanced capture kinetics: Modeling investigation of mass transfer mechanisms. *Chem. Eng. J.* **2022,** *428*, 131603.
108. Rim, G.; Feric, T. G.; Moore, T.; Park, A.-H. A., Solvent Impregnated Polymers Loaded with Liquid-Like Nanoparticle Organic Hybrid Materials for Enhanced Kinetics of Direct Air Capture and Point Source CO2 Capture. *Advanced Functional Materials* **2021,** *31* (21), 2010047.
109. Rim, G.; Wang, D.; Bardiya, V.; Stylianou, K.; Smit, B.; Lee, D.; Park, A.-H. In *Towards sustainable energy and materials: CO2 capture using novel nanoscale hybrid particulate systems*, CFB 2021- Proceedings of the 13th International Conference on Fluidized Bed Technology, GLAB Reactor and Fluidization Technologies: 2021; pp 18-25.
110. Lee, W. R.; Hwang, S. Y.; Ryu, D. W.; Lim, K. S.; Han, S. S.; Moon, D.; Choi, J.; Hong, C. S., Diamine-functionalized metal–organic framework: exceptionally high CO 2 capacities from ambient air and flue gas, ultrafast CO 2 uptake rate, and adsorption mechanism. *Energy Environ. Sci.* **2014,** *7* (2), 744-751.
111. Fracaroli, A. M.; Furukawa, H.; Suzuki, M.; Dodd, M.; Okajima, S.; Gándara, F.; Reimer, J. A.; Yaghi, O. M., Metal–Organic Frameworks with Precisely Designed Interior for Carbon Dioxide Capture in the Presence of Water. *J. Am. Chem. Soc.* **2014,** *136* (25), 8863-8866.
112. McDonald, T. M.; Mason, J. A.; Kong, X.; Bloch, E. D.; Gygi, D.; Dani, A.; Crocellà, V.; Giordanino, F.; Odoh, S. O.; Drisdell, W. S.; Vlaisavljevich, B.; Dzubak, A. L.; Poloni, R.; Schnell, S. K.; Planas, N.; Lee, K.; Pascal, T.; Wan, L. F.; Prendergast, D.; Neaton, J. B.; Smit, B.; Kortright, J. B.; Gagliardi, L.; Bordiga, S.; Reimer, J. A.; Long, J. R., Cooperative insertion of CO2 in diamine-appended metal-organic frameworks. *Nature* **2015,** *519* (7543), 303-308.
113. Liao, P.-Q.; Chen, H.; Zhou, D.-D.; Liu, S.-Y.; He, C.-T.; Rui, Z.; Ji, H.; Zhang, J.-P.; Chen, X.-M., Monodentate hydroxide as a super strong yet reversible active site for CO2 capture from high-humidity flue gas. *Energy Environ. Sci.* **2015,** *8* (3), 1011-1016.
114. Yu, W.; Gao, M.; Rim, G.; Feric, T. G.; Rivers, M. L.; Alahmed, A.; Jamal, A.; Alissa Park, A.-H., Novel in-capsule synthesis of metal–organic framework for innovative carbon dioxide capture system. *Green Energy & Environment* **2021**.

115. Wang, D.; Yu, W.; Gao, M.; Liu, K.; Wang, T.; Park, A.; Lin, Q. In *A 3D microfluidic device for carbon capture microcapsules production*, 2018 IEEE Micro Electro Mechanical Systems (MEMS), 21-25 Jan. 2018; 2018; pp 1193-1196.
116. Gores, S.; Scheffler, M.; Graichen, V., Greenhouse gas emission trends and projections in Europe 2012. Tracking progress towards Kyoto and 2020 targets. **2012**.
117. Lackner, K.; Ziock, H.-J.; Grimes, P. *Carbon dioxide extraction from air: Is it an option?*; Los Alamos National Lab., NM (US): 1999.
118. Zeman, F. S.; Lackner, K. S., Capturing carbon dioxide directly from the atmosphere. *World Resource Review* **2004,** *16* (2), 157-172.
119. Keith, D. W.; Ha-Duong, M.; Stolaroff, J. K., Climate Strategy with $CO_2$ Capture from the Air. *Climatic Change* **2006,** *74* (1), 17-45.
120. Baciocchi, R.; Storti, G.; Mazzotti, M., Process design and energy requirements for the capture of carbon dioxide from air. *Chemical Engineering and Processing: Process Intensification* **2006,** *45* (12), 1047-1058.
121. Nikulshina, V.; Hirsch, D.; Mazzotti, M.; Steinfeld, A., $CO_2$ capture from air and co-production of H2 via the Ca(OH)2–CaCO3 cycle using concentrated solar power–Thermodynamic analysis. *Energy* **2006,** *31* (12), 1715-1725.
122. Weaver, A. J.; Zickfeld, K.; Montenegro, A.; Eby, M., Long term climate implications of 2050 emission reduction targets. *Geophys. Res. Lett.* **2007,** *34* (19).
123. Zeman, F., Energy and Material Balance of $CO_2$ Capture from Ambient Air. *Environ. Sci. Technol.* **2007,** *41* (21), 7558-7563.
124. Zeman, F., Experimental results for capturing $CO_2$ from the atmosphere. *AIChE J.* **2008,** *54* (5), 1396-1399.
125. Nikulshina, V.; Ayesa, N.; Gálvez, M. E.; Steinfeld, A., Feasibility of Na-based thermochemical cycles for the capture of $CO_2$ from air—Thermodynamic and thermogravimetric analyses. *Chem. Eng. J.* **2008,** *140* (1), 62-70.
126. Stolaroff, J. K.; Keith, D. W.; Lowry, G. V., Carbon Dioxide Capture from Atmospheric Air Using Sodium Hydroxide Spray. *Environ. Sci. Technol.* **2008,** *42* (8), 2728-2735.
127. Keith, D. W., Why Capture $CO_2$ from the Atmosphere? *Science* **2009,** *325* (5948), 1654-1655.
128. Lackner, K. S., Capture of Carbon Dioxide from Ambient Air. *Eur Phys J Spec Top* **2009,** *176* (1), 93-106.
129. Sherman, S. R., Nuclear powered $CO_2$ capture from the atmosphere. *Environ. Prog. Sustain. Energy* **2009,** *28* (1), 52-59.
130. Nikulshina, V.; Gebald, C.; Steinfeld, A., $CO_2$ capture from atmospheric air via consecutive CaO-carbonation and CaCO3-calcination cycles in a fluidized-bed solar reactor. *Chem. Eng. J.* **2009,** *146* (2), 244-248.
131. Chichilnisky, G.; Eisenberger, P., How air capture could help to promote a Copenhagen solution. *Nature* **2009,** *459*, 1053.
132. Dessler, A., Energy for air capture. *Nature Geoscience* **2009,** *2*, 811.
133. Lackner, K. S.; Brennan, S., Envisioning carbon capture and storage: expanded possibilities due to air capture, leakage insurance, and C-14 monitoring. *Climatic Change* **2009,** *96* (3), 357-378.
134. Pielke Jr, R., Air capture update. *Nature Geoscience* **2009,** *2*, 811.
135. Pielke, R. A., An idealized assessment of the economics of air capture of carbon dioxide in mitigation policy. *Environ. Sci. Policy* **2009,** *12* (3), 216-225.
136. Eisenberger, P. M.; Cohen, R. W.; Chichilnisky, G.; Eisenberger, N. M.; Chance, R. R.; Jones, C. W., Global Warming and Carbon-Negative Technology: Prospects for a Lower-Cost Route to a Lower-Risk Atmosphere. *Energy & Environment* **2009,** *20* (6), 973-984.

137.	McDonald, T. M.; Lee, W. R.; Mason, J. A.; Wiers, B. M.; Hong, C. S.; Long, J. R., Capture of Carbon Dioxide from Air and Flue Gas in the Alkylamine-Appended Metal–Organic Framework mmen-Mg2(dobpdc). *J. Am. Chem. Soc.* **2012,** *134* (16), 7056-7065.
138.	van der Giesen, C.; Meinrenken, C. J.; Kleijn, R.; Sprecher, B.; Lackner, K. S.; Kramer, G. J., A Life Cycle Assessment Case Study of Coal-Fired Electricity Generation with Humidity Swing Direct Air Capture of $CO_2$ versus MEA-Based Postcombustion Capture. *Environ. Sci. Technol.* **2017,** *51* (2), 1024-1034.
139.	Davis, S. J.; Lewis, N. S.; Shaner, M.; Aggarwal, S.; Arent, D.; Azevedo, I. L.; Benson, S. M.; Bradley, T.; Brouwer, J.; Chiang, Y.-M.; Clack, C. T. M.; Cohen, A.; Doig, S.; Edmonds, J.; Fennell, P.; Field, C. B.; Hannegan, B.; Hodge, B.-M.; Hoffert, M. I.; Ingersoll, E.; Jaramillo, P.; Lackner, K. S.; Mach, K. J.; Mastrandrea, M.; Ogden, J.; Peterson, P. F.; Sanchez, D. L.; Sperling, D.; Stagner, J.; Trancik, J. E.; Yang, C.-J.; Caldeira, K., Net-zero emissions energy systems. *Science* **2018,** *360* (6396), eaas9793.
140.	National Academies of Sciences, E.; Medicine, Negative emissions technologies and reliable sequestration: a research agenda. *Negative emissions technologies and reliable sequestration: a research agenda.* **2018**.
141.	*Greenhouse gas removal*. Royal Academy of Engineering: 2018.
142.	Haszeldine, R. S.; Flude, S.; Johnson, G.; Scott, V., Negative emissions technologies and carbon capture and storage to achieve the Paris Agreement commitments. *Philosophical Transactions of the Royal Society A: Mathematical, Physical and Engineering Sciences* **2018,** *376* (2119), 20160447.
143.	Keith, D. W.; Holmes, G.; St. Angelo, D.; Heidel, K., A Process for Capturing $CO_2$ from the Atmosphere. *Joule* **2018,** *2* (8), 1573-1594.
144.	Masson-Delmotte, V., P. Zhai, H.-O. Pörtner, D. Roberts, J. Skea, P.R. Shukla, A. Pirani, W. Moufouma-Okia, C. Péan, R. Pidcock, S. Connors, J.B.R. Matthews, Y. Chen, X. Zhou, M.I. Gomis, E. Lonnoy, Maycock, M. Tignor, and T. Waterfield (eds.), *IPCC* **2018**, 32.
145.	Fasihi, M.; Efimova, O.; Breyer, C., Techno-economic assessment of $CO_2$ direct air capture plants. *Journal of Cleaner Production* **2019,** *224*, 957-980.
146.	de Jonge, M. M. J.; Daemen, J.; Loriaux, J. M.; Steinmann, Z. J. N.; Huijbregts, M. A. J., Life cycle carbon efficiency of Direct Air Capture systems with strong hydroxide sorbents. *Int. J. Greenh. Gas Control.* **2019,** *80*, 25-31.
147.	Hu, Z.; Wang, Y.; Shah, B. B.; Zhao, D., $CO_2$ Capture in Metal–Organic Framework Adsorbents: An Engineering Perspective. *Advanced Sustainable Systems* **2019,** *3* (1), 1800080.
148.	McQueen, N.; Gomes, K. V.; McCormick, C.; Blumanthal, K.; Pisciotta, M.; Wilcox, J., A review of direct air capture (DAC): scaling up commercial technologies and innovating for the future. *Progress in Energy* **2021,** *3* (3), 032001.
149.	Shi, X.; Xiao, H.; Azarabadi, H.; Song, J.; Wu, X.; Chen, X.; Lackner, K. S., Sorbents for the Direct Capture of CO2 from Ambient Air. *Angew. Chem. Int. Ed.* **2020,** *59* (18), 6984-7006.
150.	United Nations. Adoption of the Paris Agreement. *United Nations:* **2015**, p 31.
151.	Shekhah, O.; Belmabkhout, Y.; Chen, Z.; Guillerm, V.; Cairns, A.; Adil, K.; Eddaoudi, M., Made-to-order metal-organic frameworks for trace carbon dioxide removal and air capture. *Nature communications* **2014,** *5*.
152.	Oschatz, M.; Antonietti, M., A search for selectivity to enable $CO_2$ capture with porous adsorbents. *Energy Environ. Sci.* **2018,** *11* (1), 57-70.
153.	Presser, V.; McDonough, J.; Yeon, S.-H.; Gogotsi, Y., Effect of pore size on carbon dioxide sorption by carbide derived carbon. *Energy Environ. Sci.* **2011,** *4* (8), 3059-3066.
154.	Sevilla, M.; Parra, J. B.; Fuertes, A. B., Assessment of the Role of Micropore Size and N-Doping in $CO_2$ Capture by Porous Carbons. *ACS Appl. Mater. Interfaces* **2013,** *5* (13), 6360-6368.
155.	Silvestre-Albero, A.; Rico-Francés, S.; Rodríguez-Reinoso, F.; Kern, A. M.; Klumpp, M.; Etzold, B. J. M.; Silvestre-Albero, J., High selectivity of TiC-CDC for $CO_2$/N2 separation. *Carbon* **2013,** *59*, 221-228.

156. Kolle, J. M.; Fayaz, M.; Sayari, A., Understanding the Effect of Water on CO2 Adsorption. *Chem. Rev.* **2021,** *121* (13), 7280-7345.
157. Wilson, S. M. W.; Tezel, F. H., Direct Dry Air Capture of CO2 Using VTSA with Faujasite Zeolites. *Ind. Eng. Chem. Res.* **2020,** *59* (18), 8783-8794.
158. Keith, D. W.; Ha-Duong, M.; Stolaroff, J. K., Climate strategy with CO2 capture from the air. *Climatic Change* **2006,** *74* (1-3), 17-45.
159. Rodríguez-Mosqueda, R.; Bramer, E. A.; Roestenberg, T.; Brem, G., Parametrical Study on $CO_2$ Capture from Ambient Air Using Hydrated K2CO3 Supported on an Activated Carbon Honeycomb. *Ind. Eng. Chem. Res.* **2018,** *57* (10), 3628-3638.
160. Beaudoin, G.; Nowamooz, A.; Assima, G. P.; Lechat, K.; Gras, A.; Entezari, A.; Kandji, E. H. B.; Awoh, A.-S.; Horswill, M.; Turcotte, S.; Larachi, F.; Dupuis, C.; Molson, J.; Lemieux, J.-M.; Maldague, X.; Plante, B.; Bussière, B.; Constantin, M.; Duchesne, J.; Therrien, R.; Fortier, R., Passive Mineral Carbonation of Mg-rich Mine Wastes by Atmospheric $CO_2$. *Energy Procedia* **2017,** *114*, 6083-6086.
161. Zhao, S.; Ma, L.; Yang, J.; Zheng, D.; Liu, H.; Yang, J., Mechanism of $CO_2$ Capture Technology Based on the Phosphogypsum Reduction Thermal Decomposition Process. *Energy Fuels* **2017,** *31* (9), 9824-9832.
162. Nikulshina, V.; Gálvez, M. E.; Steinfeld, A., Kinetic analysis of the carbonation reactions for the capture of $CO_2$ from air via the Ca(OH)2–CaCO3–CaO solar thermochemical cycle. *Chem. Eng. J.* **2007,** *129* (1), 75-83.
163. Nikulshina, V.; Steinfeld, A., $CO_2$ capture from air via CaO-carbonation using a solar-driven fluidized bed reactor—Effect of temperature and water vapor concentration. *Chem. Eng. J.* **2009,** *155* (3), 867-873.
164. Yanase, I.; Onozawa, S.; Ohashi, Y.; Takeuchi, T., $CO_2$ capture from ambient air by β-NaFeO2 in the presence of water vapor at 25–100 °C. *Powder Technology* **2019,** *348*, 43-50.
165. Chaikittisilp, W.; Kim, H.-J.; Jones, C. W., Mesoporous alumina-supported amines as potential steam-stable adsorbents for capturing $CO_2$ from simulated flue gas and ambient air. *Energy Fuels* **2011,** *25* (11), 5528-5537.
166. Chaikittisilp, W.; Lunn, J. D.; Shantz, D. F.; Jones, C. W., Poly (L‐lysine) Brush–Mesoporous Silica Hybrid Material as a Biomolecule‐Based Adsorbent for $CO_2$ Capture from Simulated Flue Gas and Air. *Chemistry–A European Journal* **2011,** *17* (38), 10556-10561.
167. Kuwahara, Y.; Kang, D. Y.; Copeland, J. R.; Bollini, P.; Sievers, C.; Kamegawa, T.; Yamashita, H.; Jones, C. W., Enhanced $CO_2$ Adsorption over Polymeric Amines Supported on Heteroatom‐Incorporated SBA‐15 Silica: Impact of Heteroatom Type and Loading on Sorbent Structure and Adsorption Performance. *Chem. Eur. J.* **2012,** *18* (52), 16649-16664.
168. Choi, S.; Watanabe, T.; Bae, T.-H.; Sholl, D. S.; Jones, C. W., Modification of the Mg/DOBDC MOF with amines to enhance $CO_2$ adsorption from ultradilute gases. *J. Phys. Chem. Lett.* **2012,** *3* (9), 1136-1141.
169. Didas, S. A.; Kulkarni, A. R.; Sholl, D. S.; Jones, C. W., Role of amine structure on carbon dioxide adsorption from ultradilute gas streams such as ambient air. *ChemSusChem* **2012,** *5* (10), 2058-2064.
170. Sakwa-Novak, M. A.; Jones, C. W., Steam Induced Structural Changes of a Poly (ethylenimine) Impregnated γ-Alumina Sorbent for $CO_2$ Extraction from Ambient Air. *ACS Appl. Mater. Interfaces* **2014,** *6* (12), 9245-9255.
171. Didas, S. A.; Choi, S.; Chaikittisilp, W.; Jones, C. W., Amine–oxide hybrid materials for $CO_2$ capture from ambient air. *Acc. Chem. Res.* **2015,** *48* (10), 2680-2687.
172. Holewinski, A.; Sakwa-Novak, M. A.; Jones, C. W., Linking $CO_2$ Sorption Performance to Polymer Morphology in Aminopolymer/Silica Composites through Neutron Scattering. *J. Am. Chem. Soc.* **2015,** *137* (36), 11749-11759.
173. Seipp, C. A.; Williams, N. J.; Kidder, M. K.; Custelcean, R., $CO_2$ Capture from Ambient Air by Crystallization with a Guanidine Sorbent. *Angew. Chem. Int. Ed.* **2017,** *56* (4), 1042-1045.

174. Brethomé, F. M.; Williams, N. J.; Seipp, C. A.; Kidder, M. K.; Custelcean, R., Direct air capture of CO2 via aqueous-phase absorption and crystalline-phase release using concentrated solar power. *Nature Energy* **2018**, 1.
175. Custelcean, R., Direct air capture of CO2 via crystal engineering. *Chemical Science* **2021**.
176. Custelcean, R.; Garrabrant, K. A.; Agullo, P.; Williams, N. J., Direct air capture of CO2 with aqueous peptides and crystalline guanidines. *Cell Reports Physical Science* **2021,** *2* (4), 100385.
177. Custelcean, R.; Williams, N. J.; Garrabrant, K. A.; Agullo, P.; Brethomé, F. M.; Martin, H. J.; Kidder, M. K., Direct Air Capture of CO2 with Aqueous Amino Acids and Solid Bis-iminoguanidines (BIGs). *Ind. Eng. Chem. Res.* **2019,** *58* (51), 23338-23346.
178. Kasturi, A.; Gabitto, J.; Tsouris, C.; Custelcean, R., Carbon dioxide capture with aqueous amino acids: Mechanistic study of amino acid regeneration by guanidine crystallization and process intensification. *Separation and Purification Technology* **2021,** *271*, 118839.
179. Kasturi, A.; Gabitto, J. F.; Custelcean, R.; Tsouris, C., A Process Intensification Approach for CO2 Absorption Using Amino Acid Solutions and a Guanidine Compound. *Energies* **2021,** *14* (18), 5821.
180. Liu, M.; Custelcean, R.; Seifert, S.; Kuzmenko, I.; Gadikota, G., Hybrid Absorption–Crystallization Strategies for the Direct Air Capture of CO2 Using Phase-Changing Guanidium Bases: Insights from in Operando X-ray Scattering and Infrared Spectroscopy Measurements. *Ind. Eng. Chem. Res.* **2020,** *59* (47), 20953-20959.
181. Williams, N. J.; Seipp, C. A.; Brethomé, F. M.; Ma, Y.-Z.; Ivanov, A. S.; Bryantsev, V. S.; Kidder, M. K.; Martin, H. J.; Holguin, E.; Garrabrant, K. A.; Custelcean, R., CO2 Capture via Crystalline Hydrogen-Bonded Bicarbonate Dimers. *Chem* **2019,** *5* (3), 719-730.
182. Wang, T.; Lackner, K. S.; Wright, A., Moisture Swing Sorbent for Carbon Dioxide Capture from Ambient Air. *Environ. Sci. Technol.* **2011,** *45* (15), 6670-6675.
183. Wang, T.; Lackner, K. S.; Wright, A. B., Moisture-swing sorption for carbon dioxide capture from ambient air: a thermodynamic analysis. *Physical Chemistry Chemical Physics* **2013,** *15* (2), 504-514.
184. Shi, X.; Li, Q.; Wang, T.; Lackner, K. S., Kinetic analysis of an anion exchange absorbent for $CO_2$ capture from ambient air. *PLoS One* **2017,** *12* (6), e0179828.
185. Song, J.; Liu, J.; Zhao, W.; Chen, Y.; Xiao, H.; Shi, X.; Liu, Y.; Chen, X., Quaternized Chitosan/PVA Aerogels for Reversible $CO_2$ Capture from Ambient Air. *Ind. Eng. Chem. Res.* **2018,** *57* (14), 4941-4948.
186. Wang, X.; Song, J.; Chen, Y.; Xiao, H.; Shi, X.; Liu, Y.; Zhu, L.; He, Y.-L.; Chen, X., CO2 Absorption over Ion Exchange Resins: The Effect of Amine Functional Groups and Microporous Structures. *Ind. Eng. Chem. Res.* **2020,** *59* (38), 16507-16515.
187. Wang, T.; Hou, C.; Ge, K.; Lackner, K. S.; Shi, X.; Liu, J.; Fang, M.; Luo, Z., Spontaneous Cooling Absorption of $CO_2$ by a Polymeric Ionic Liquid for Direct Air Capture. *J. Phys. Chem. Lett.* **2017,** *8* (17), 3986-3990.
188. Wang, T.; Wang, X.; Hou, C.; Liu, J., Quaternary functionalized mesoporous adsorbents for ultra-high kinetics of CO2 capture from air. *Sci. Rep.* **2020,** *10* (1), 21429.
189. Hou, C.; Wu, Y.; Wang, T.; Wang, X.; Gao, X., Preparation of Quaternized Bamboo Cellulose and Its Implication in Direct Air Capture of CO2. *Energy Fuels* **2019,** *33* (3), 1745-1752.
190. Han, Y.; Zhu, L.; Yao, Y.; Shi, X.; Zhang, Y.; Xiao, H.; Chen, X., Strong bases behave as weak bases in nanoscale chemical environments: implication in humidity-swing CO2 air capture. *Phys. Chem. Chem. Phys.* **2021,** *23* (27), 14811-14817.
191. Shi, X.; Xiao, H.; Chen, X.; Lackner, K. S., The Effect of Moisture on the Hydrolysis of Basic Salts. *Chem. Eur. J.* **2016,** *22* (51), 18326-18330.
192. Shi, X.; Xiao, H.; Kanamori, K.; Yonezu, A.; Lackner, K. S.; Chen, X., Moisture-Driven CO2 Sorbents. *Joule* **2020,** *4* (8), 1823-1837.
193. Song, J.; Zhu, L.; Shi, X.; Liu, Y.; Xiao, H.; Chen, X., Moisture Swing Ion-Exchange Resin-PO4 Sorbent for Reversible CO2 Capture from Ambient Air. *Energy Fuels* **2019,** *33* (7), 6562-6567.

194.	Armstrong, M.; Shi, X.; Shan, B.; Lackner, K.; Mu, B., Rapid $CO_2$ capture from ambient air by sorbent-containing porous electrospun fibers made with the solvothermal polymer additive removal technique. *AIChE J.* **2019,** *65* (1), 214-220.
195.	Shi, X.; Xiao, H.; Liao, X.; Armstrong, M.; Chen, X.; Lackner, K. S., Humidity effect on ion behaviors of moisture-driven $CO_2$ sorbents. *Int. J. Chem. Phys.* **2018,** *149* (16), 164708.
196.	Parzuchowski, P. G.; Świderska, A.; Roguszewska, M.; Rolińska, K.; Wołosz, D., Moisture- and Temperature-Responsive Polyglycerol-Based Carbon Dioxide Sorbents—The Insight into the Absorption Mechanism for the Hydrophilic Polymer. *Energy Fuels* **2020,** *34* (10), 12822-12832.
197.	Voskian, S.; Hatton, T. A., Faradaic electro-swing reactive adsorption for CO2 capture. *Energy Environ. Sci.* **2019,** *12* (12), 3530-3547.
198.	Kang, J. S.; Kim, S.; Hatton, T. A., Redox-responsive sorbents and mediators for electrochemically based CO2 capture. *Current Opinion in Green and Sustainable Chemistry* **2021,** *31*, 100504.
199.	Singh, P.; Rheinhardt, J. H.; Olson, J. Z.; Tarakeshwar, P.; Mujica, V.; Buttry, D. A., Electrochemical Capture and Release of Carbon Dioxide Using a Disulfide–Thiocarbonate Redox Cycle. *J. Am. Chem. Soc.* **2017,** *139* (3), 1033-1036.
200.	Apaydin, D. H.; Gora, M.; Portenkirchner, E.; Oppelt, K. T.; Neugebauer, H.; Jakesova, M.; Głowacki, E. D.; Kunze-Liebhäuser, J.; Zagorska, M.; Mieczkowski, J.; Sariftci, N. S., Electrochemical Capture and Release of $CO_2$ in Aqueous Electrolytes Using an Organic Semiconductor Electrode. *ACS Appl. Mater. Interfaces* **2017,** *9* (15), 12919-12923.
201.	Rheinhardt, J. H.; Singh, P.; Tarakeshwar, P.; Buttry, D. A., Electrochemical Capture and Release of Carbon Dioxide. *ACS Energy Letters* **2017,** *2* (2), 454-461.
202.	Su, X.; Bromberg, L.; Martis, V.; Simeon, F.; Huq, A.; Hatton, T. A., Postsynthetic Functionalization of Mg-MOF-74 with Tetraethylenepentamine: Structural Characterization and Enhanced CO2 Adsorption. *ACS Appl. Mater. Interfaces* **2017,** *9* (12), 11299-11306.
203.	Kim, E. J.; Siegelman, R. L.; Jiang, H. Z. H.; Forse, A. C.; Lee, J.-H.; Martell, J. D.; Milner, P. J.; Falkowski, J. M.; Neaton, J. B.; Reimer, J. A.; Weston, S. C.; Long, J. R., Cooperative carbon capture and steam regeneration with tetraamine-appended metal-organic frameworks. *Science* **2020,** *369* (6502), 392-396.
204.	Bien, C. E.; Liu, Q.; Wade, C. R., Assessing the Role of Metal Identity on CO2 Adsorption in MOFs Containing M–OH Functional Groups. *Chem. Mater.* **2020,** *32* (1), 489-497.
205.	Wang, B.; Lin, R.-B.; Zhang, Z.; Xiang, S.; Chen, B., Hydrogen-Bonded Organic Frameworks as a Tunable Platform for Functional Materials. *J. Am. Chem. Soc.* **2020,** *142* (34), 14399-14416.
206.	Zhou, Y.; Zhang, J.; Wang, L.; Cui, X.; Liu, X.; Wong, S. S.; An, H.; Yan, N.; Xie, J.; Yu, C.; Zhang, P.; Du, Y.; Xi, S.; Zheng, L.; Cao, X.; Wu, Y.; Wang, Y.; Wang, C.; Wen, H.; Chen, L.; Xing, H.; Wang, J., Self-assembled iron-containing mordenite monolith for carbon dioxide sieving. *Science* **2021,** *373* (6552), 315-320.
207.	Qazvini, O. T.; Babarao, R.; Telfer, S. G., Selective capture of carbon dioxide from hydrocarbons using a metal-organic framework. *Nature Communications* **2021,** *12* (1), 197.
208.	Trickett, C. A.; Helal, A.; Al-Maythalony, B. A.; Yamani, Z. H.; Cordova, K. E.; Yaghi, O. M., The chemistry of metal–organic frameworks for CO2 capture, regeneration and conversion. *Nature Reviews Materials* **2017,** *2* (8), 17045.
209.	Hu, X.; Liu, L.; Luo, X.; Xiao, G.; Shiko, E.; Zhang, R.; Fan, X.; Zhou, Y.; Liu, Y.; Zeng, Z.; Li, C. e., A review of N-functionalized solid adsorbents for post-combustion CO2 capture. *Applied Energy* **2020,** *260*, 114244.
210.	Kumar, A.; Madden, D. G.; Lusi, M.; Chen, K. J.; Daniels, E. A.; Curtin, T.; Perry, J. J.; Zaworotko, M. J., Direct air capture of $CO_2$ by physisorbent materials. *Angew. Chem. Int. Ed.* **2015,** *54* (48), 14372-14377.

211. Madden, D. G.; Scott, H. S.; Kumar, A.; Chen, K.-J.; Sanii, R.; Bajpai, A.; Lusi, M.; Curtin, T.; Perry, J. J.; Zaworotko, M. J., Flue-gas and direct-air capture of $CO_2$ by porous metal–organic materials. *Phil. Trans. R. Soc. A* **2017,** *375* (2084), 20160025.
212. Findley, J. M.; Sholl, D. S., Computational Screening of MOFs and Zeolites for Direct Air Capture of Carbon Dioxide under Humid Conditions. *J. Phys. Chem. C* **2021,** *125* (44), 24630-24639.
213. Shekhah, O.; Belmabkhout, Y.; Chen, Z.; Guillerm, V.; Cairns, A.; Adil, K.; Eddaoudi, M., Made-to-order metal-organic frameworks for trace carbon dioxide removal and air capture. *Nature Communications* **2014,** *5* (1), 4228.
214. Mukherjee, S.; Sikdar, N.; O'Nolan, D.; Franz, D. M.; Gascón, V.; Kumar, A.; Kumar, N.; Scott, H. S.; Madden, D. G.; Kruger, P. E.; Space, B.; Zaworotko, M. J., Trace $CO_2$ capture by an ultramicroporous physisorbent with low water affinity. *Science Advances* **2019,** *5* (11), eaax9171.
215. Caplow, M., Kinetics of carbamate formation and breakdown. *J. Am. Chem. Soc.* **1968,** *90* (24), 6795-6803.
216. Laddha, S. S.; Danckwerts, P. V., Reaction of $CO_2$ with ethanolamines: kinetics from gas-absorption. *Chem. Eng. Sci.* **1981,** *36* (3), 479-482.
217. Pinto, M. L.; Mafra, L.; Guil, J. M.; Pires, J.; Rocha, J., Adsorption and activation of $CO_2$ by amine-modified nanoporous materials studied by solid-state NMR and $13CO_2$ adsorption. *Chem. Mater.* **2011,** *23* (6), 1387-1395.
218. Yu, J.; Chuang, S. S. C., The Role of Water in $CO_2$ Capture by Amine. *Ind. Eng. Chem. Res.* **2017,** *56* (21), 6337-6347.
219. D'Alessandro, D. M.; Smit, B.; Long, J. R., Carbon dioxide capture: prospects for new materials. *Angew. Chem. Int. Ed.* **2010,** *49* (35), 6058-6082.
220. Yang, Z.-Z.; He, L.-N.; Zhao, Y.-N.; Li, B.; Yu, B., $CO_2$ capture and activation by superbase/polyethylene glycol and its subsequent conversion. *Energy Environ. Sci.* **2011,** *4* (10), 3971-3975.
221. Lee, J. J.; Chen, C.-H.; Shimon, D.; Hayes, S. E.; Sievers, C.; Jones, C. W., Effect of Humidity on the $CO_2$ Adsorption of Tertiary Amine Grafted SBA-15. *J. Phys. Chem. C* **2017,** *121* (42), 23480-23487.
222. Dutcher, B.; Fan, M.; Russell, A. G., Amine-Based $CO_2$ Capture Technology Development from the Beginning of 2013—A Review. *ACS Appl. Mater. Interfaces* **2015,** *7* (4), 2137-2148.
223. Goeppert, A.; Meth, S.; Prakash, G. S.; Olah, G. A., Nanostructured silica as a support for regenerable high-capacity organoamine-based $CO_2$ sorbents. *Energy Environ. Sci.* **2010,** *3* (12), 1949-1960.
224. Xu, X.; Song, C.; Andresen, J. M.; Miller, B. G.; Scaroni, A. W., Novel polyethylenimine-modified mesoporous molecular sieve of MCM-41 type as high-capacity adsorbent for $CO_2$ capture. *Energy Fuels* **2002,** *16* (6), 1463-1469.
225. Xu, X.; Song, C.; Andresen, J. M.; Miller, B. G.; Scaroni, A. W., Preparation and characterization of novel $CO_2$ "molecular basket" adsorbents based on polymer-modified mesoporous molecular sieve MCM-41. *Microporous and mesoporous materials* **2003,** *62* (1), 29-45.
226. Xu, X.; Song, C.; Miller, B. G.; Scaroni, A. W., Adsorption separation of carbon dioxide from flue gas of natural gas-fired boiler by a novel nanoporous "molecular basket" adsorbent. *Fuel Processing Technology* **2005,** *86* (14), 1457-1472.
227. Franchi, R. S.; Harlick, P. J. E.; Sayari, A., Applications of Pore-Expanded Mesoporous Silica. 2. Development of a High-Capacity, Water-Tolerant Adsorbent for $CO_2$. *Ind. Eng. Chem. Res.* **2005,** *44* (21), 8007-8013.
228. Yue, M. B.; Sun, L. B.; Cao, Y.; Wang, Y.; Wang, Z. J.; Zhu, J. H., Efficient $CO_2$ Capturer Derived from As‐Synthesized MCM‐41 Modified with Amine. *Chemistry–A European Journal* **2008,** *14* (11), 3442-3451.
229. Brilman, D.; Veneman, R., Capturing atmospheric $CO_2$ using supported amine sorbents. *Energy Procedia* **2013,** *37*, 6070-6078.

230. Choi, S.; Gray, M. L.; Jones, C. W., Amine-Tethered Solid Adsorbents Coupling High Adsorption Capacity and Regenerability for $CO_2$ Capture From Ambient Air. *ChemSusChem* **2011,** *4* (5), 628-635.
231. Sanz-Pérez, E. S.; Arencibia, A.; Calleja, G.; Sanz, R., Tuning the textural properties of HMS mesoporous silica. Functionalization towards $CO_2$ adsorption. *Microporous and Mesoporous Materials* **2018,** *260*, 235-244.
232. Belmabkhout, Y.; Serna-Guerrero, R.; Sayari, A., Adsorption of $CO_2$-Containing Gas Mixtures over Amine-Bearing Pore-Expanded MCM-41 Silica: Application for Gas Purification. *Ind. Eng. Chem. Res.* **2010,** *49* (1), 359-365.
233. Sanz, R.; Calleja, G.; Arencibia, A.; Sanz-Pérez, E. S., $CO_2$ capture with pore-expanded MCM-41 silica modified with amino groups by double functionalization. *Microporous and Mesoporous Materials* **2015,** *209*, 165-171.
234. Franchi, R.; Harlick, P. J. E.; Sayari, A., A high capacity, water tolerant adsorbent for $CO_2$: diethanolamine supported on pore-expanded MCM-41. *Studies in Surface Science and Catalysis* **2005,** *156*, 879-886.
235. Filburn, T.; Helble, J. J.; Weiss, R. A., Development of Supported Ethanolamines and Modified Ethanolamines for $CO_2$ Capture. *Ind. Eng. Chem. Res.* **2005,** *44* (5), 1542-1546.
236. Plaza, M. G.; Pevida, C.; Arenillas, A.; Rubiera, F.; Pis, J. J., $CO_2$ capture by adsorption with nitrogen enriched carbons. *Fuel* **2007,** *86* (14), 2204-2212.
237. Gibson, J. A. A.; Gromov, A. V.; Brandani, S.; Campbell, E. E. B., The effect of pore structure on the $CO_2$ adsorption efficiency of polyamine impregnated porous carbons. *Microporous and Mesoporous Materials* **2015,** *208*, 129-139.
238. Sujan, A.; Pang, S. H.; Zhu, G.; Jones, C. W.; Lively, R. P., Direct $CO_2$ capture from air using poly(ethyleneimine)-loaded polymer/silica fiber sorbents. *ACS Sustainable Chemistry & Engineering* **2019**.
239. Chen, C.; Son, W.-J.; You, K.-S.; Ahn, J.-W.; Ahn, W.-S., Carbon dioxide capture using amine-impregnated HMS having textural mesoporosity. *Chem. Eng. J.* **2010,** *161* (1), 46-52.
240. Ma, X.; Wang, X.; Song, C., "Molecular Basket" Sorbents for Separation of CO2 and H2S from Various Gas Streams. *J. Am. Chem. Soc.* **2009,** *131* (16), 5777-5783.
241. Klinthong, W.; Huang, C.-H.; Tan, C.-S., One-Pot Synthesis and Pelletizing of Polyethylenimine-Containing Mesoporous Silica Powders for CO2 Capture. *Ind. Eng. Chem. Res.* **2016,** *55* (22), 6481-6491.
242. Heydari-Gorji, A.; Yang, Y.; Sayari, A., Effect of the Pore Length on CO2 Adsorption over Amine-Modified Mesoporous Silicas. *Energy Fuels* **2011,** *25* (9), 4206-4210.
243. Darunte, L. A.; Sen, T.; Bhawanani, C.; Walton, K. S.; Sholl, D. S.; Realff, M. J.; Jones, C. W., Moving Beyond Adsorption Capacity in Design of Adsorbents for $CO_2$ Capture from Ultradilute Feeds: Kinetics of $CO_2$ Adsorption in Materials with Stepped Isotherms. *Ind. Eng. Chem. Res.* **2019,** *58* (1), 366-377.
244. Liao, P.-Q.; Chen, X.-W.; Liu, S.-Y.; Li, X.-Y.; Xu, Y.-T.; Tang, M.; Rui, Z.; Ji, H.; Zhang, J.-P.; Chen, X.-M., Putting an ultrahigh concentration of amine groups into a metal–organic framework for $CO_2$ capture at low pressures. *Chemical Science* **2016,** *7* (10), 6528-6533.
245. Planas, N.; Dzubak, A. L.; Poloni, R.; Lin, L.-C.; McManus, A.; McDonald, T. M.; Neaton, J. B.; Long, J. R.; Smit, B.; Gagliardi, L., The Mechanism of Carbon Dioxide Adsorption in an Alkylamine-Functionalized Metal–Organic Framework. *J. Am. Chem. Soc.* **2013,** *135* (20), 7402-7405.